\documentclass[reprint,twocolumn,prx,amsmath,amssymb,superscriptaddress,floatfix]{revtex4-2}

\usepackage{amsmath}
\usepackage{amssymb}
\usepackage{mathrsfs}
\usepackage{graphicx}
\usepackage{dsfont}
\usepackage{color}
\usepackage{amsfonts}
\usepackage{subfigure}
\usepackage{enumitem}
\usepackage{lipsum}
\usepackage[normalem]{ulem}

\begin{document}

\title{Individual- and pair-based models of epidemic spreading: master equations and analysis of their forecasting capabilities}

\author{Federico Malizia}
\affiliation{Department of Physics and Astronomy,  University of Catania, 95125 Catania, Italy}
\author{Luca Gallo}
\affiliation{Department of Physics and Astronomy,  University of Catania, 95125 Catania, Italy}
\affiliation{INFN Sezione di Catania, Via S. Sofia, 64, 95125 Catania, Italy}
\author{Mattia Frasca}
\affiliation{Department of Electrical, Electronics and Computer Science Engineering, University of Catania, 95125 Catania, Italy}
\author{Vito Latora}
\affiliation{Department of Physics and Astronomy,  University of Catania, 95125 Catania, Italy}
\affiliation{INFN Sezione di Catania, Via S. Sofia, 64, 95125 Catania, Italy}
\affiliation{School of Mathematical Sciences, Queen Mary University of London, London E1 4NS, UK}
\affiliation{Complexity Science Hub Vienna, A-1080 Vienna, Austria}
\author{Giovanni Russo}
\affiliation{Department of Mathematics and Computer Science, University of Catania, 95125 Catania, Italy}
    
\date{\today}

\begin{abstract}
Epidemic models are crucial to understand how an infectious disease spreads in a population, and to devise the best containment strategies. Compartmental models can provide a fine-grained description of the evolution of an epidemic when microscopic information on the network of contacts among individuals is available. However, coarser-grained descriptions prove also to be useful in many aspects. 
They allow to derive closed expressions for key parameters, such as the basic reproduction number, to understand the relationship between the model parameters, and also to derive fast and reliable predictions of macroscopic observables for a disease outbreak. The so-called \emph{population models} can be developed at different levels 
of coarse-graining, so that it is crucial to determine: \emph{ i)} to which extent and how the existing correlations in the contact network have to be included in these models, and \emph{ ii)} what is their impact on the model ability to reproduce and predict the time evolution of the populations at the various stage of the disease. In this work, we address these questions through a systematic analysis of two discrete-time SEAIR (Susceptible-Exposed-Asymptomatic-Infected-Recovered) population models: the first one developed assuming statistical independence at the level of individuals, and the other one assuming independence at the level of pairs.
We provide a detailed derivation and analysis of both models, focusing on their capability to reproduce an epidemic process on different synthetic networks, and then comparing their predictions under scenarios of increasing complexity.
We find that, although both models can fit the time evolution of the compartment populations obtained through microscopic simulations, the epidemic parameters adopted by the individual-based model for this purpose may significantly differ from those of the microscopic simulations. However, pair-based model provides not only more reliable predictions of the dynamical evolution of the variables, but also a good estimation of the epidemic parameters. The difference between the two models is even more evident in the particularly challenging scenario when one or more variables are not measurable, and therefore are not available for model tuning. This occurs for instance with asymptomatic infectious individuals in the case of COVID-19, an issue that has become  extremely relevant during the recent pandemic. 
Under these conditions, the pairwise model again proves to perform much better than the individual-based representation, provided that it is fed with adequate information which, for instance, to be collected, may require a more detailed contact tracing. Overall, our results thus hallmark the importance of acquiring the proper empirical data to fully unfold the potentialities of models incorporating more sophisticated assumptions on the correlations among nodes in the contact network.

\end{abstract}
\maketitle

\section{Introduction}

The last decades have witnessed the emergence of new infectious diseases and the resurgence of old ones \cite{mcmichael2004environmental, fauci2012perpetual, morens2013emerging}. Examples include the outbreak of Ebola in West Africa \cite{coltart2017ebola}, the epidemic of Zika virus in North and South America \cite{fauci2016zika}, and the global spreading of respiratory diseases such as influenza A(H1N1) \cite{novel2009emergence} and, more recently, COVID-19 \cite{estrada2020covid,li2020early}, all of which have risen international public health concern. In this context, mathematical modeling of disease spreading has played a fundamental role in the understanding, control and prevention of epidemic outbreaks, guiding the policy-making processes through quantitative analyses \cite{keeling2011modeling,pastor2015epidemic,kiss2017mathematics}. As the spreading of an infectious disease within a population is heavily affected by the precise patterns of contacts among individuals and by their mobility habits, 
complex networks, which allow to represent the intricate structure of human interactions \cite{newman2003structure,boccaletti2006complex,latora2017complex}, are 
a fundamental tool for analyzing the dynamics of an epidemic. 
A variety of techniques exists to model the dynamics of a disease spreading on a network \cite{pastor2015epidemic,kiss2017mathematics}, ranging from macroscopic compartmental models based on ordinary differential equations \cite{kermack1927contribution,anderson1992infectious}, to more sophisticated data-driven agent-based microscopic simulations \cite{eubank2004modelling, longini2005containing, ferguson2005strategies} and metapopulation structured models \cite{colizza2008epidemic,ajelli2010comparing}. 
In particular, deterministic representations of compartmental models can be formulated, aiming at describing the epidemic process in terms of the temporal evolution of the probabilities that a node is in a given state. To this purpose, one can adopt either a top-down approach, which consists in defining all the possible configurations of the network and the mechanisms ruling transitions from one network state to another, or a bottom-up approach, focusing instead on the state of single nodes \cite{kiss2017mathematics}. The latter approach leads to a hierarchy of coupled differential/master equations. Indeed, the dynamics of a node typically depends on the state of the node itself and of its neighbors, thus on the state of pairs of nodes, which in turns depends on triples, in a hierarchy of dynamical correlations and dependencies.

Such hierarchies are very common in kinetic and statistical physics. For example, classical derivation of the Boltzmann equation of rarefied gas dynamics is based on the so-called BBGKY hierarchy, formally obtained by the integration of the Liouville equation of a system of $N$ particles undergoing binary collisions: the evolution equation for the $k$-particle distribution function depends on the $(k+1)$-particle distribution function \cite{Cercignani}. 
As another example, in Extended Thermodynamics, the evolution equations of moments of order $k$ of the distribution function depend on the moments of order $k+1$ \cite{muller2013rational}. 
In all cases, to be of practical use the hierarchies have to be truncated by introducing some approximation. For instance, in the case of the Boltzmann equation, the hierarchy is closed at the level of single particle distribution function by the so-called {\em Stosszahlansatz} and {\em propagation of chaos}, while in extended thermodynamics, the closure is usually based on the so-called maximum entropy principle \cite{Cercignani,muller2013rational}.

In the context of mathematical epidemiology, similar bottom-up approaches have been adopted to develop individual and pair-based continuous-time models on networks, and applied, for instance, to the SIR process in the homogeneous \cite{rogers2011maximum} and heterogeneous \cite{bell2021beyond} mean-field approximation. Individual and pair-based approximations for the SIR process have also been analyzed in the discrete-time case where, unlike the standard continuous-time BBGKY-type hierarchy, the master equations describing the dynamics of the system are expressed in terms of joint probabilities, whose order is governed by the network structure itself via the degrees of individual nodes \cite{frasca2016discrete}, thus resulting in a richer although more complicated mathematical structure.
A relevant question concerning these kinds of hierarchies is at which order one needs to truncate them in order to get accurate predictions about an epidemic process. Indeed, while it has been shown how to close the system of equations at various orders \cite{         ,house2009motif}, it is less clear how the chosen approximation affects the features of the model, and, more specifically, its ability to forecast the dynamical evolution of the state variables. In particular, a systematic analysis of the reliability of the predictions obtained by different hierarchy closures and under different hypotheses on the quantity and quality of available data,   is still lacking. Indeed, the problem of data availability is of utmost importance in the context of the COVID-19 pandemic, where the epidemic process has been proven difficult to characterize and the mechanisms at work have not yet been fully unveiled \cite{vespignani2020modelling,roda2020difficult,gallo2020lack}.
In this paper we address the issues above focusing on the SEAIR model, a compartmental model with five compartments.  
This is a generalization of the SIR model  accounting for two critical features of infectious diseases such as  the COVID-19, namely the existence of a latency period and the presence of asymptomatic carriers. While being rather simple, the SEAIR model allows to study the crucial case of two infectious population, namely the symptomatic and the asymptomatic, with the latter that, given the difficulties in its measurement, had a significant role in the COVID-19 pandemic \cite{he2020temporal,li2020early,rothe2020transmission,lavezzo2020suppression}. In more detail, first we derive the discrete-time master equations for the \emph{SEAIR model} with two different orders of approximation, closing the system both at the level of individuals (individual-based population models) and at the level of pairs (pair-based population models), and deriving the corresponding set of master equations. We then compare these two approximations, analyzing to which extent they can capture the temporal evolution of the SEAIR epidemic process. To do so, we  consider scenarios of increasing complexity, with the hypotheses on the amount of available data gradually becoming more strict from case to case. 

\medskip
The paper is organized as follows. First, we describe the microscopic mechanisms underlying the epidemic process and discuss the mathematical modeling approaches to describe it in Sec.~\ref{section:model_introduction}. We then derive an individual-based and a pair-based version of the SEAIR model in Sec.~\ref{section:models}. Assuming to know the epidemiological characteristics, in Sec.~\ref{section:epidemic_evolution} we compare the ability of the models in reproducing the dynamics of an outbreak. In Sec.~\ref{section:fitting} we analyse the predictive capability of the models when parts of the information on the epidemic are not available. Finally, we summarize and discuss the main findings of the work in Sec.~\ref{section:conclusion}

\section{Modeling an epidemic process\label{section:model_introduction}}

In compartmental models the population is partitioned into several states, namely the \emph{compartments,} representing the different stages of the disease course. For instance, in the SIR model, which is one of the simplest compartmental models, an individual can either be susceptible (S), infected (I), or recovered/removed (R). To model the spreading of a disease 
within a population, the most important step is to characterize the processes governing the transitions of individuals from one disease stage, to another i.e. from a compartment to another \cite{pastor2015epidemic}. In the SIR model, the contagion is defined by two fundamental mechanisms. First, we have a two-body nonlinear process, representing the infection of a susceptible individual (S) by an infected one (I), which acts as a mediator of the transition. Second, we have a one-body linear process, describing the recovery (R) of an infected individual (I). The transitions from one compartment to another can be formally expressed with the following kinetic equations   
\begin{equation}
\begin{array}{rcl}
     S + I & \overset{\beta}{\rightarrow} & I + I \\
     I & \overset{\mu}{\rightarrow} & R,
\end{array}
\label{eq:SIR_transitions}
\end{equation}
where $\beta$ and $\mu$ represent, in the case of a continuous-time model, the transition rates for the infection and recovery processes, while they can be interpreted as transition probabilities in the case of a discrete-time model.
These are two tunable control parameters of the model, which can be fixed based on the previous knowledge we have of the disease under study.  
For instance, the parameter regulating the recovery process, i.e. $\mu$, can be set from the knowledge of the typical infectious period of the disease, i.e. the average time during which an individual remains infectious before recovering. In a discrete time model, $\mu$ represents the probability that an infected individual recovers in a time unit $\Delta t$, typically one day. The expected number of time units for recovery is therefore $1/\mu$. The infectious period for influenza, for example, commonly lasts about 2-8 days, while, for smallpox, it can last over 20 days \cite{kretzschmar2009mathematical}, resulting in a smaller value of $\mu$ for smallpox, compared to the value characterizing influenza.

From compartmental models the basic reproduction number, $\mathcal{R}_0$, of an infectious disease \cite{diekmann1990definition} can be evaluated. This parameter, which represents the average number of secondary infections produced by an infectious individual at an early stage of an epidemic, is of utmost relevance in mathematical biology, as it determines whether an infection can spread across the population. Quantitatively, when $\mathcal{R}_0<1$, i.e. when the number of new infections per infectious individual is less than one, the infection is not able to spread. Otherwise, when $\mathcal{R}_0>1$, an epidemic outbreak can occur \cite{dietz1993estimation}. In the simple case of the SIR model, $\mathcal{R}_0$ is given by the ratio between the infection and the recovery rates/probabilities, i.e. $\mathcal{R}_0 = \beta/\mu$.

\begin{figure}[t]
    \centering
    \includegraphics[width=\linewidth]{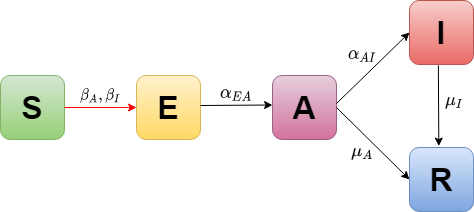}
    \caption{Graphical representation of the SEAIR model considered in this work. The model has five compartments, representing 
    the susceptible (S), exposed (E), asymptomatic infectious (A), symptomatic infectious  (I),  and  recovered/removed  (R) state. Allowed transitions from one state to another are indicated by arrows, 
    so that the flow diagram of the model can be represented as a 
    directed graph with an associated adjacency matrix $\mathscr{A}_I$ (see Section \ref{sec:pair_model}). Transition probabilities are governed by the six parameters reported close to the arrows and in Eq.~\ref{eq:SEAIR_transitions}. Edges in black represent linear transitions between states, while the edge in red refers to the nonlinear one.}
    \label{fig:SEAIR_model}
\end{figure}

In this paper we consider a discrete-time compartmental model with five compartments, in which,  
at any time, an individual can be either susceptible (S), exposed to the disease but still unable to spread it (E), asymptomatic infectious (A), symptomatic infectious (I), or recovered/removed (R). We refer to this compartmental model as the SEAIR model \cite{arenas2020modeling,basnarkov2021seair}. 
This is a generalization of the SIR model, as it incorporates two additional states, i.e.\ the E and A compartments, which describe two essential aspects of several infectious diseases, namely the existence of a latency period and the presence of asymptomatic carriers \cite{nelson2014infectious}.
In particular, the presence of infectious individuals with no 
symptoms of the disease (state A) 
may play a critical role in an epidemic outbreak, as it has been observed, for instance, in the recent COVID-19 pandemic \cite{he2020temporal,li2020early,rothe2020transmission,lavezzo2020suppression}.
In the SEAIR model the progress of an individual through the disease stages is determined by the following kinetic equations
\begin{equation} 
\begin{array}{rcl}
     S + A & \overset{\beta_A}{\rightarrow} & E + A  \\
     S + I & \overset{\beta_I}{\rightarrow} & E + I \\
     E & \overset{\alpha_{EA}}{\rightarrow} & A \\
     A & \overset{\alpha_{AI}}{\rightarrow} & I \\
     A & \overset{\mu_A}{\rightarrow} & R \\
     I & \overset{\mu_I}{\rightarrow} & R.
\end{array}
\label{eq:SEAIR_transitions}
\end{equation}
that are graphically illustrated in the flow diagram of Fig.~\ref{fig:SEAIR_model}.
At variance with the SIR model, the SEAIR model accounts for two different types of infection mechanisms, as a susceptible individual (in state S) may be infected upon a contact with either an asymptomatic (in state A) or a symptomatic (in state I) infectious  individual. These processes are described by the first two kinetic equations, and are governed by two infection probabilities, $\beta_A$ and $\beta_I$ respectively, which may take  different values. 

These probabilities depend on several factors.  For relatively short time step $\Delta t$ of the discrete-time model, we can imagine that the probability $\beta_A$ that a susceptible individual gets infected by an asymptomatic one in such a time step can be expressed as 
$\beta_A = \nu_A \times p_A \times \Delta t$, where $\nu_A$ represents the rate of encounters of an asymptomatic individual with other individuals, and $p_A$ the probability that one such encounter transmits the infection. The contact frequency depends on the awareness of the health condition: susceptible individuals tend to avoid contacts with symptomatic people more than with other susceptible or asymptomatic individuals. Furthermore, the probability that one encounter transmits the infection, $p_A$, depends both on the intrinsic transmissibility of the disease, and on the preventive measures adopted, such as social distancing, use of masks and disinfectant, and so on \footnote{The number $n = \nu_A\Delta t$ represents the average number of encounters. If $p_A$ is the probability of getting infected after one encounter, so that the probability {\em of not getting infected after one encounter\/} is $1-p_A$, the probability of getting infected after $n$ encounters is $1-(1-p_A)^n$. We shall make frequent use of this formula in the evaluation of the transition probabilities of our model}. For example, in the case of COVID-19, it has been shown that the absence of symptoms, such as coughing and sneezing, applies a smaller a-priori transmissibility of the disease, and therefore a small value of $p_I$. It is therefore reasonable to assume $\beta_A < \beta_I$ \cite{he2020relative,byambasuren2020estimating}. 
However, as asymptomatic individuals can be unaware about their condition, there are high chances that they will maintain their habitual social behavior. 
Therefore, as asymptomatic individuals can spread the virus without knowing it, they constitute an important public-health risk. 
Indeed, it is reasonable to consider $\beta_A>\beta_I$, assuming that there is a higher chance of dangerous contacts with subjects with no symptoms, thus incorporating such increased contact rates in the value of the transmission probability $p_A$, and consequently on $\beta_A$ \cite{giordano2020modelling}.
Hence, the possibility to tune separately the 
values of the two parameters $\beta_I$ and $\beta_A$, and to consider both the  regime with $\beta_A>\beta_I$ and the one with $\beta_A<\beta_I$, is an important aspect of the SEAIR model. 
The other important feature, which makes the SEAIR different from the SIR model, is the addition of the $E$ compartment accounting for the presence of individuals exposed to the disease but still unable to spread it. 
Notice, from the flow diagram in  Fig.~\ref{fig:SEAIR_model}, that in the SEIAR model the newly infected individuals first move to the $E$ state, meaning that they are not immediately able to spread the disease. Before becoming infectious, the exposed individuals undergo a latency period, after which they are able to spread the disease as asymptomatic carriers. The transition from $E$ to $A$ is governed by the probability $\alpha_{EA}$, which is the parameter controlling the average duration of the latency period $\tau = \Delta t / \alpha_{EA}$. Asymptomatic individuals can eventually develop symptoms, and this is taken into account by the transition from state A to state 
I, which takes place with probability $\alpha_{AI}$. 
Subsequently, the individuals recover, or are removed, either with or without having previously shown symptoms, with two different probabilities, 
$\mu_I$ and $\mu_A$ respectively. In conclusion, the SEAIR model is based on five compartments and is ruled by six different tunable parameters $\beta_A, \beta_I, \alpha_{EA}, \alpha_{AI}, \mu_A, \mu_I$.

\medskip
As the disease can only spread through encounters between infectious and susceptible individuals, the progress of an epidemic outbreak ultimately depends on the patterns of contacts among the individual themselves. 
These are strongly ruled not only by social habits, but also by the geographical distributions of the individuals and by the way these distributions change in time. Humans travel across a hierarchy of characteristic spatial scales, such as neighbourhoods, cities and countries, so that their highly heterogeneous mobility influences the way in which they interact 
\cite{brockmann2006scaling,gonzalez2008understanding,song2010modelling,arcaute20, alessandretti20}.
Therefore, a crucial aspect of any realistic modeling of epidemic spreading is how to model the contact patterns within a population. 
Large-scale agent-based simulations 
\cite{ferguson2005strategies}
and structured metapopulation models based on data-driven mobility schemes at the interpopulation level \cite{Chinazzi2020} are usually complemented 
by models amenable to mathematical analysis
\cite{gomez2018critical}
that capture the influence of human behaviour and the existence of complex social structures. In this context, complex networks \cite{latora2017complex} 
have revealed particularly useful as they allow to represent the physical contact patterns that result from real movements of individuals between specific locations
\cite{eubank2004modelling}, and also to investigate 
in a controlled way how the social structure of a population affects the evolution and outcome of an epidemic.

Here, we will focus on network modelling of disease spreading. Namely, we will implement the stochastic processes that describe the infectious disease on an undirected graph $\mathcal{G} = (\mathcal{V},\mathcal{E})$, with $N=|\mathcal{V}|$ nodes and $K = |\mathcal{E}|$ links. 
In the following we will indicate as $G= \{ g_{ij}\}$ (with 
$g_{ij}=1$ if node $i$ and $j$ are connected, while $g_{ij}=0$ otherwise) the $N \times N$ adjacency matrix of such graph.
In this framework, each node $i \in \mathcal{V}$ of the graph represents an individual of the population, and can be in one of the five 
different states of SEAIR model. 
Each edge $(i,j) \in \mathcal{E}$ represents a contact along which the infection can take place. 
Nodes change states according to  Eqs.~(\ref{eq:SEAIR_transitions}), 
where the first two kinetic equations are implemented over the links of the graph. 
Different graph topologies 
will be studied in this article. For simplicity, however, the graphs considered here will always be fixed in time, which means that the pattern of contacts is assumed not to change during the evolution of the disease. This is, however, a strong assumption, as humans tend to react to a disease by avoiding contacts with 
infected individuals. This leads to 
a rewiring of links depending on the states of the nodes that affects the dynamics of the disease, which in turn influences the rewiring process. Temporal networks and time-varying graphs are an important subfield of network science \cite{Holme_rev12,holme2013temporal,masuda_guide_temp_net}, and simple epidemic models, such as the SIS, have been studied on temporal networks \cite{buscarino2008,tang2010sw}
and on adaptive networks, i.e. on networks whose structure is coevolving with the disease \cite{gross_adaptive}. Although the SEAIR model can be straightforwardly implemented on a time-varying graph, this is beyond the scope of this paper.

The dynamics of the disease spreading on a network can be studied in several different ways \cite{pastor2015epidemic,kiss2017mathematics}. Focusing on probabilistic methods, one can decide to investigate a system 
numerically, performing extensive stochastic simulations of the epidemic process on the network, or, alternatively, one can develop a deterministic representation of the process, writing the evolution equations for the probability that the network is in a given state.
In general, the deterministic equations can be derived at the level of each single node or at the population (mean-field) level. The first approach requires to consider a large number of evolution equations that can be impractical and computationally costly to integrate. Furthermore, empirical data on both the contact network and the state of the nodes may be limited or unavailable. For instance, the information on the state of the individuals is usually provided at a coarser grain, i.e. at a \emph{population level} \cite{giordano2020modelling, pcdata}. Finally, from a microscopic stochastic simulation on a large network, it is more difficult to understand how the emerging collective behavior is related to the underlying parameters, which could be better interpreted by a mesoscopic coarse-grained modeling. For these reasons, it is common to assume that the individuals are homogeneously mixed and interact with each other completely at random. Under this hypothesis, each node of the network can be considered statistically equivalent to any other, which permits to describe the system at a population level, drastically reducing the number of evolution equations.

In Section \ref{section:models}, we develop two population-level models for the contact-based SEAIR epidemic process described in Eqs.~(\ref{eq:SEAIR_transitions}). First, we assume statistical independence at the level of nodes, deriving a model we refer to as the \emph{individual-based SEAIR model}. Next, we assume  statistical independence at the level of node pairs, crucially incorporating the dynamical correlations within the network in a model we refer to as the \emph{pair-based} (or \emph{pairwise}) \emph{SEAIR model}.

\section{Population-level models \label{section:models}}

To begin with, here we discuss the dynamical variables characterizing the individual-based and the pairwise SEAIR models. For the individual-based model, the description of the system is carried out at the level of single individuals, i.e. the nodes, while for the pairwise, one has to express the dynamics at the level of pairs of individuals, i.e. the edges. Mathematically, this means to write a set of master equations governing the temporal evolution of either the probability that a node is in a given compartment, in the case of an individual-based model, or the probability that an edge is in a certain state, in the case of a pairwise model. Let us denote as
$\Omega$ the set of possible compartments, i.e., $\Omega=\{S,E,A,I,R\}$ in the SEAIR model. Adopting a standard notation \cite{kiss2017mathematics}, we indicate 
as $\langle X_i \rangle_t$ the probability that a node $i$ belongs to the compartment $X \in \Omega$ at time $t$. 
For the sake of convenience, let us 
denote as U the state of a node which is 
not infectious, namely a node that 
belongs to either S, E or R. We denote with $\langle U_i \rangle_t$ the probability that a node $i$ is the state 
U at time $t$. Such a probability can be evaluated as $\langle U_i \rangle_t = \langle S_i \rangle_t + \langle E_i \rangle_t + \langle R_i \rangle_t$.
Under the homogeneous mixing hypothesis, each node of the network is assumed to be statistically equivalent to any other, meaning that $\langle X_i \rangle_t = \langle X_j \rangle_t$ for $i,j=1,\ldots N$. Hence, for the population level model, we can drop the node index, as we have
\begin{equation}
\langle X_i\rangle_t = \langle X\rangle_t = \frac{[X]_t}{N}, \quad 
\forall i=1,\ldots N, \quad \forall X \in \Omega,
\label{eq:individual_variables}
\end{equation}
where $[X]_t$ represents the expected number of individuals in the compartment $X$ at time $t$, and $N$ is the number of individuals in the population. Here, $\langle X\rangle_t$ is the probability that a generic node of the network is found in the state $X$ at time $t$, and represents the fundamental variable of the individual-based model. 
Analogously, the probability that a generic node is in a non-infectious state U at time $t$ is $\langle U \rangle_t = \langle S \rangle_t + \langle E \rangle_t + \langle R \rangle_t$.

Similarly, we denote as $\langle X_i Y_j \rangle$ the probability that the edge $(i,j)$ is in state $(X_i,Y_j)$. Again, in a mean-field approximation, we can drop the indices as we have 
\begin{equation}
\langle X_iY_j\rangle_t = \langle XY\rangle_t = \frac{[XY]_t}{2K}, 
\forall i,j=1,\ldots N,
\quad X,Y\in\Omega
\label{eq:pair_variables}
\end{equation}
where $[XY]_t$ represents the expected number of pairs in state $(X,Y)$ at time $t$, $K$ is the number of edges in the network, and $\langle XY\rangle_t$ represents the probability that a generic link of the network is found in the state $(X,Y)$ at time $t$. The quantities $\langle XY\rangle_t$, with $X,Y \in \Omega$, will be the fundamental variables of the pair-based model. 
Note that from Eq.~(\ref{eq:pair_variables}) it follows that $\langle XY \rangle_t = \langle YX \rangle_t$. The individual (node) state  probabilities in any compartment, $\langle X \rangle_t$, can be obtained as the marginal probabilities of the pair (edge) state probabilities as
\begin{equation}
\label{eq:marginal}
\langle X\rangle_t = \sum_{Y \in \Omega} \langle XY \rangle_t.
\end{equation}

In the most general case, the notation $\langle X_{i_{1}}Y_{i_{2}} \ldots Z_{i_{n}} \rangle_t$ denotes the probability that a given connected subgraph induced by nodes $i_1$,$i_2$,$\ldots$,$i_n$ 
is found in the state $(X_{i_{1}}Y_{i_{2}} \ldots Z_{i_{n}})$ at time $t$. Once again, under the homogeneous mixing hypothesis, we can write such a quantity as $\langle X Y \ldots Z\rangle_t$.

\bigskip
Now, in order to develop an {\em individual-based model} for the SEAIR dynamics, one has to write a set of equation describing how the individual state probabilities $\langle X \rangle_t$ in Eq.~(\ref{eq:individual_variables}) evolve in time. To do so, however, it is necessary to introduce an hypothesis of statistical independence at the level of the individuals. Indeed, the infection processes described in  Eq.~(\ref{eq:SEAIR_transitions}) require to consider the joint probability that an individual is susceptible and that one of its contacts in the network is infectious, either asymptomatic or symptomatic, i.e. the probabilities $\langle SA\rangle_t$ and $\langle SI\rangle_t$, respectively. Therefore, to write a set of equations involving only the individual state probabilities, i.e. a closed-form expression, one has to express these higher-order probabilities in terms of the quantities $\langle X\rangle_t$. Mathematically, this is done by assuming statistical independence at the level of the nodes, i.e. writing  $\langle XY\rangle_t \approx \langle X\rangle_t \langle Y\rangle_t$. 
However, this assumption overlooks the impact of dynamical correlations that indeed exist within the contact network, e.g. infected nodes are more likely to be in contact with other infected nodes \cite{kiss2017mathematics}. 

To take these correlations into account, one can develop a {\em pair-based model} of the SEAIR dynamics, describing how the pair state probabilities in Eq.~(\ref{eq:pair_variables}) evolve in time \cite{pastor2015epidemic}. As we will see thereafter, even the pairwise model will require to express some high-order joint probabilities in terms of lower-order probabilities, in this case $\langle X\rangle_t$ and $\langle XY\rangle_t$.

In the rest of this section, we formulate the master equations for both the individual-based and the pairwise models, and discuss under which hypotheses a closed set of equations can be obtained.

\subsection{Individual-based SEAIR model}
\label{sub:ib}
Here we derive the individual-based population-level model of the SEAIR epidemic process.
%
As previously discussed, under the hypothesis of homogeneous mixing, we can write a set of evolution equations at the population level for the probability $\langle X \rangle_t$ that a generic node in the network belongs to a given compartment $X$ at time $t$. Following the transition diagram in Fig.~\ref{fig:SEAIR_model}, we can write the master equations (also known as rate equations) of the system as
\begin{equation}
\label{eq:individual}
\begin{array}{lll}
    \langle S \rangle_{t+1} &=& \langle S \rangle_{t} - \Pi_{S \rightarrow E}
    \\
    \langle E \rangle_{t+1} &=& \langle E \rangle_{t} + \Pi_{S \rightarrow E} - \Pi_{E \rightarrow A}
    \\
    \langle A \rangle_{t+1} &=& \langle A \rangle_{t} + \Pi_{E \rightarrow A} - \Pi_{A \rightarrow I} - \Pi_{A \rightarrow R}
    \\
    \langle I \rangle_{t+1} &=& \langle I \rangle_{t} + \Pi_{A \rightarrow I} - \Pi_{I \rightarrow R}
    \\
    \langle R \rangle_{t+1} &=&
    \langle R \rangle_{t} + \Pi_{I \rightarrow R} + \Pi_{A \rightarrow R},
\end{array}
\end{equation}
where the term $\Pi_{X \rightarrow Y}$ denotes the probability that at a generic node there is a transition from state $X$ at time $t$ to state $Y$ at time $t+1$.

As we have seen, in the epidemic process defined by 
the kinetic equations in Eq.~(\ref{eq:SEAIR_transitions}) we identify two classes of transitions. On the one hand, we have the class of two-body nonlinear processes, for which the transition of a given node from one compartment to another needs to be mediated by the interactions with 
other nodes, as the node can only be infected by one of its neighbors. On the other hand, we have the class of one-body linear processes, for which the transitions of an individual from a state to another does not depend on the state of its contacts. Consistently, we distinguish two kinds of transition probabilities, with the 
one-body processes being described by linear probability terms, whereas the two-body processes give rise to nonlinear terms.  

Let us first consider an example of linear transition. The probability of moving from state $I$ to state $R$, which corresponds to the recovery process of a symptomatic individual, can be written as
\begin{equation}
    \Pi_{I\rightarrow R} = \langle I \rangle_t\mu_I ,
    \label{eq:individual_linear_trans}
\end{equation}
where $\mu_I$ represents the recovery probability for the symptomatic individuals. As the recovery of a symptomatic individual does not depend on the state of the other individuals, the transition probability is only determined by the probability of being in state $I$ at time $t$, i.e. the term is linear in $\langle I \rangle_t$.
Similarly, we can write the remaining linear transition probabilities as
\begin{equation}
    \begin{array}{lll}
    \Pi_{E \rightarrow A} &=& \langle E \rangle_{t}\alpha_{EA}
    \\
    \Pi_{A \rightarrow I} &=& \langle A\rangle_{t}\alpha_{AI}
    \\
    \Pi_{A \rightarrow R} &=&\langle A\rangle_{t}\mu_A,
    \end{array}
    \label{eq:all_individual_linear_trans}
\end{equation}
where $\alpha_{EA}$, $\alpha_{AI}$ and $\mu_A$ represent the probability that an individual becomes infectious, develops symptoms and recovers while asymptomatic, respectively.

As an example of nonlinear term, let us now consider the transition probability regulating the infection of a susceptible individual, i.e.\ the probability $\Pi_{S\rightarrow E}$. At variance with the recovery processes, an infection can only occur if a susceptible individual interacts with an infectious one. In other words, the transition probability of a node from state $S$ to state $E$ depends on the state of its nearest neighbors. Therefore, in order to express this term at a population level, we should first develop a node-level description, and thereafter introduce the homogeneous mixing hypothesis. 
\begin{figure}[t]
    \centering
    \includegraphics[width=0.7\linewidth]{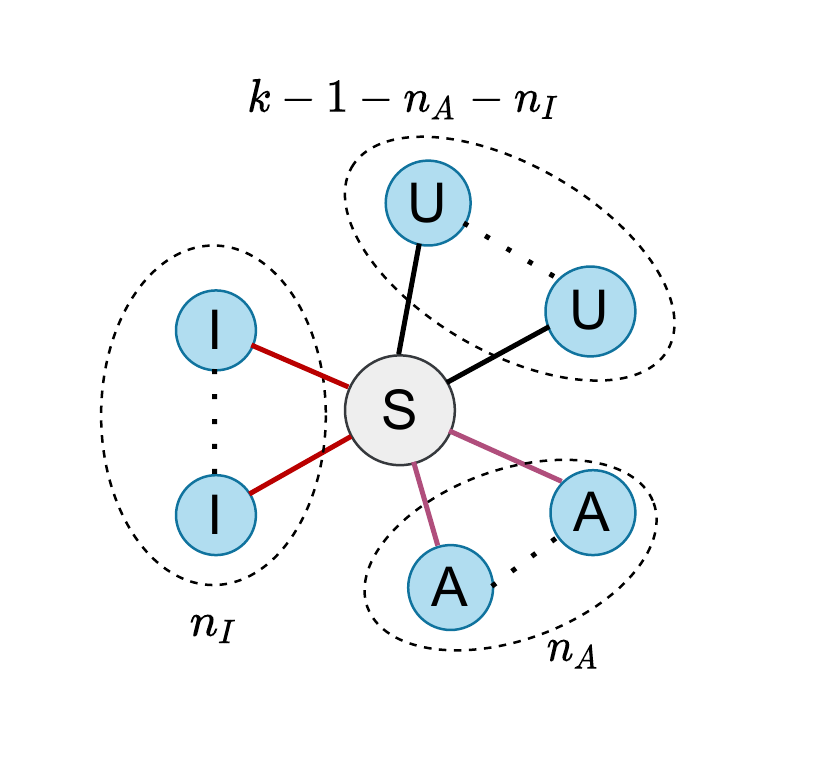}
    \caption{Graphical representation of a node $i$ with its $k$ neighbors. Here, $n_A$ indicates the number of neighbors of $i$ in state A, and, similarly, $n_I$ those in state I.}
    \label{fig:sottografo_individual}
\end{figure}
To do this, let us denote the probability that a node $i$ moves from state $S$ to state $E$ at time $t$ as $\Pi_{S_i \rightarrow E_{i}}$. 
If we consider the simple case of a node $i$ of 
degree $k_i = 1$, i.e. connected to only one other node, say $j$, we can write the transition probability as
\begin{equation}
    \begin{array}{ll}
    \Pi_{S_i \rightarrow E_i} =& \langle S_i I_{j}\rangle_t \beta_I + \langle S_i A_{j}\rangle_t \beta_A.
    \end{array}
\end{equation}
In the case where $k_i = 2$, we have instead 
\begin{equation}
    \begin{array}{ll}
    \Pi_{S_i \rightarrow E_i} =& \langle S_iI_{j_1}U_{j_2}\rangle_t [1-(1-\beta_I)]\\
    &+ \langle S_iU_{j_1}I_{j_2}\rangle_t [1-(1-\beta_I)]\\
    &+ \langle S_iI_{j_1}I_{j_2}\rangle_t [1-(1-\beta_I)^2]\\
    &+ \langle S_iA_{j_1}U_{j_2}\rangle_t [1-(1-\beta_A)]\\
    &+ \langle S_iU_{j_1}A_{j_2}\rangle_t [1-(1-\beta_A)]\\
    &+ \langle S_iA_{j_1}A_{j_2}\rangle_t [1-(1-\beta_A)^2]\\
    &+ \langle S_iI_{j_1}A_{j_2}\rangle_t [1-(1-\beta_I)(1-\beta_A)]\\
    &+ \langle S_iA_{j_1}I_{j_2}\rangle_t [1-(1-\beta_I)(1-\beta_A)].
    \end{array}
\end{equation}

In the general case when node $i$ is connected to $k$ other nodes, i.e. $k_i=k$, as shown in Fig.~\ref{fig:sottografo_individual}, we can write the transition probability as
\begin{equation}
    \begin{array}{ll}
    \Pi_{S_i \rightarrow E_i} =& \langle S_iI_{j_1}U_{j_2}\ldots U_{j_k} \rangle_t [1-(1-\beta_I)] \\
    &+ \langle S_iU_{j_1}I_{j_2}\ldots U_{j_k} \rangle_t [1-(1-\beta_I)]\\
    & +\ldots\\
    &+ \langle S_iU_{j_1}U_{j_2}\ldots I_{j_k} \rangle_t [1-(1-\beta_I)]\\
    &+ \langle S_iA_{j_1}U_{j_2}\ldots U_{j_k} \rangle_t [1-(1-\beta_A)]\\
    &+\ldots\\
    &+ \langle S_iU_{j_1}U_{j_2}\ldots A_{j_k} \rangle_t [1-(1-\beta_A)]\\
    &+ \langle S_iI_{j_1}I_{j_2}\ldots U_{j_k} \rangle_t [1-(1-\beta_I)^2]\\
    &+\ldots\\
    &+ \langle S_iI_{j_1}U_{j_2}\ldots I_{j_k} \rangle_t [1-(1-\beta_I)^2]\\
    &+ \langle S_iI_{j_1}A_{j_2}\ldots U_{j_k} \rangle_t [1-(1-\beta_I)(1-\beta_A)]\\
    &+\ldots\\
    &+ \langle S_iA_{j_1}A_{j_2}\ldots A_{j_k} \rangle_t [1-(1-\beta_A)^k],
    \end{array}
    \label{eq:transition_individual}
\end{equation}
where $\beta_A$ and $\beta_I$ are the infection probabilities for the asymptomatic and symptomatic, respectively. Each term on the right-hand side is the product of two terms, namely the joint probability that node $i$ belongs to the state $S$ and its neighborhood is in a certain state, multiplied by the conditional probability that $i$ gets infected over the next time step, given that particular state of its neighborhood.

Now, going back to Eqs.~(\ref{eq:individual}), we note that they are exact but not closed, because Eq.~(\ref{eq:transition_individual}) involves the joint probability of the $(k+1)$-uple formed by the node and its $k$ neighbors. Indeed, the evolution of such joint probability, in turn, would depend on the joint probability of larger set of nodes, giving rise to a hierarchy of coupled equations of increasing complexity. As mentioned in the Introduction, this  
type of hierarchies are very common in statistical physics. 
Examples are the BBGKY hierarchy in kinetic theory \cite{Cercignani},
and the moment hierarchy in extended thermodynamics \cite{muller2013rational}. As it would be unpractical or even impossible to deal with the full hierarchy of evolution equations, it is common to close the system by expressing the higher-order joint probabilities in terms of lower-order ones. 
In the context of epidemic models, various {\em closure methods} have been adopted to develop individual and pair based  models of continuous-time stochastic systems, both in the  
homogeneous and heterogeneous mean field approximation 
\cite{kiss2017mathematics,rogers2011maximum,bell2021beyond}. 

In this section, we close the system of equations~(\ref{eq:individual}) at the level of individual nodes by assuming statistical independence of their states. Under this hypothesis, we can approximate the $(k+1)$-body joint probabilities as:
\begin{equation}
    \langle S_i X_{j_1}Y_{j_2}\ldots Z_{j_k} \rangle_t \approx \langle S_i \rangle_t \langle X_{j_1}\rangle_t \langle Y_{j_2}\rangle_t \ldots \langle Z_{j_k} \rangle_t.
    \label{eq:individual_closure}
\end{equation}
Given the expressions in Eq.~(\ref{eq:individual_closure}) we can rewrite the joint probability terms appearing in Eq.~(\ref{eq:transition_individual}). Then, 
by considering the individuals within the populations to be homogeneously mixed and by assuming that the number of contacts of each node is fixed and equal to $k$ (i.e. $k_i=k ~\forall i$), after some algebra (see Appendix A for details), we can express the population-level transition probability from state $S$ to state $E$ as 
\begin{equation}
\label{eq:PiSEchiusaInd}
    \Pi_{S \rightarrow E} \approx \langle S \rangle_t \left[1 - (1- \beta_A \langle A\rangle_t - \beta_I \langle I\rangle_t)^k\right].
\end{equation}

To summarize, a closed-form, individual-based population-level SEAIR model is obtained by replacing in Eq.~(\ref{eq:individual}) 
the transition term $\Pi_{S \rightarrow E}$ as in Eq.~(\ref{eq:PiSEchiusaInd}), and the linear transition probabilities given by Eq.~(\ref{eq:individual_linear_trans}) and Eq.~(\ref{eq:all_individual_linear_trans}). 

To conclude, hereafter we report the value of the basic reproduction number, i.e. $\mathcal{R}_0$, which permits to determine whether the infectious disease is able to spread across the population. 
We compute it by using the next-generation matrix (NGM) approach \cite{diekmann1990definition,diekmann2010construction}, developed for discrete-time epidemic models \cite{allen2008basic} (more details on the method are reported in Appendix D). We have 
\begin{equation}
\label{eq:R0_individual}
    \mathcal{R}_0 = \dfrac{k\left(\alpha_{AI}\beta_I + \mu_I \beta_A\right)}{\mu_I\left(\alpha_{AI}+\mu_A\right)}.
\end{equation}

\subsection{Pair-based SEAIR model \label{sec:pair_model}}
\begin{figure}[t!]
    \centering
    \includegraphics[width=\linewidth]{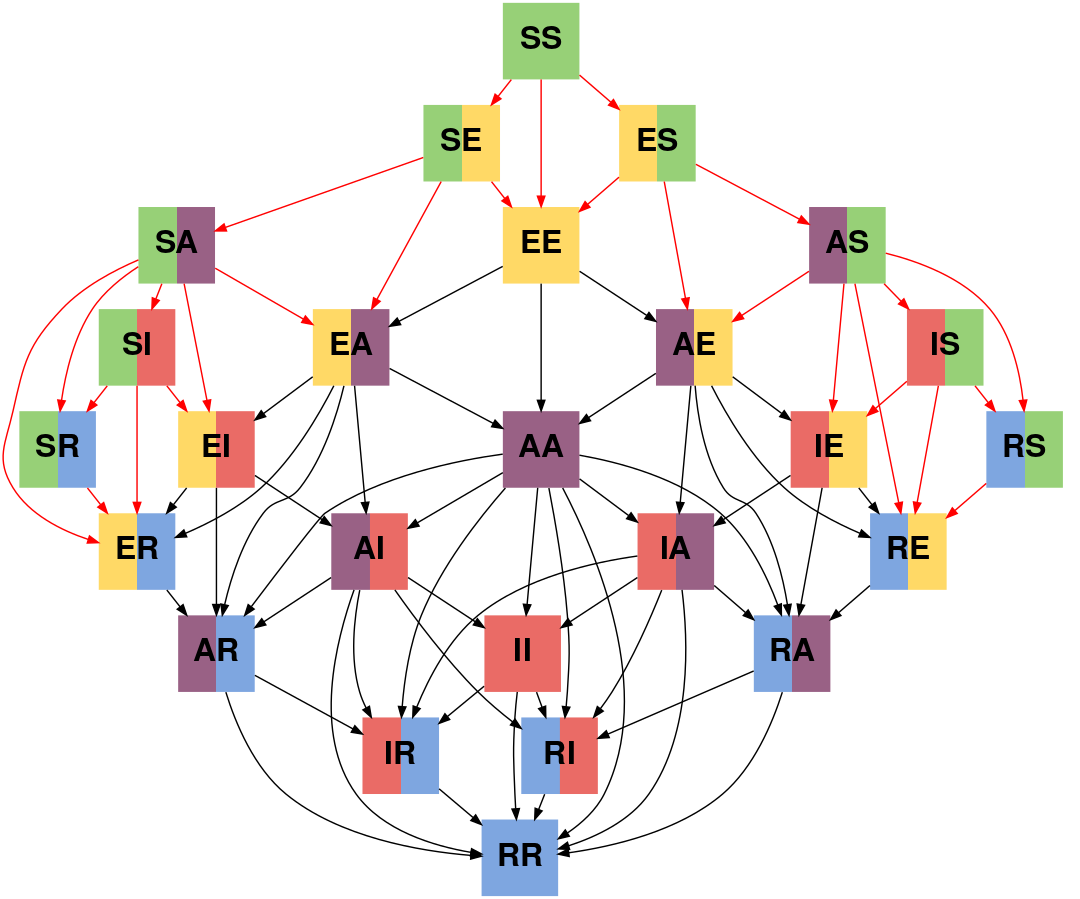}
    \caption{Graphical representation of the twenty-five different pair states (i.e. states of pairs of node) of the pairwise SEAIR model, and of all the possible transitions from one state to another described by a 
    directed flow graph with an associated adjacency matrix $\mathscr{A}_P$. Edges in black represent linear transitions between states, while edges in red refer to nonlinear ones.}
    \label{fig:pairwise_transitions}
\end{figure}

At variance with the individual-based model 
of Section \ref{sub:ib}, for which we have obtained the master equations for the individual  probabilities $\langle X \rangle_t$, in order to 
derive a pair-based (or pairwise)  population-level SEAIR model, we need to write down a closed set of equations describing the temporal evolution of the pair probabilities $\langle XY \rangle_t$. To do so, we first need to list all the possible transitions that may occur among the pair states. This conceptual step is not straightforward, as the number of ways in which a pair can move from one state to another according to Eq.~(\ref{eq:SEAIR_transitions}) is considerably larger than in the case of individual transitions. For instance, a pair in state $(S,S)$ can progress either to one of the states $(S,E)$ or $(E,S)$, when only one node of the pair is infected, or to the state $(E,E)$, when both nodes are infected at the same time. 
Let us come back for a moment to the diagram of the individual transitions shown in Fig.~\ref{fig:SEAIR_model}. We can interpret the flow diagram as a directed graph and associate to it an adjacency matrix $\mathscr{A}_I$. If we label the model compartments such that S is indexed by 1, E by 2, and so on, then we have that $(\mathscr{A}_I)_{nm}=1$ (with $n,m = 1,\ldots 5$) if there is a transition from compartment $m$ to compartment $n$, and $(\mathscr{A}_I)_{nm}=0$ otherwise. Note that, for any $n$, $(\mathscr{A}_I)_{nn}=1$, as an individual in state $n$ can remain in it, although, for the sake of clarity, the self-loops are not represented in Fig.~\ref{fig:SEAIR_model}. 
Therefore, the adjacency matrix $\mathscr{A}_I$ corresponding to the diagram in Fig.~\ref{fig:SEAIR_model}
is given by 
\begin{equation*}
    \mathscr{A}_I = 
    \left(
        \begin{array}{ccccc}
            1 & 1 & 0 & 0 & 0 \\
            0 & 1 & 1 & 0 & 0 \\
            0 & 0 & 1 & 1 & 1 \\
            0 & 0 & 0 & 1 & 1 \\
            0 & 0 & 0 & 0 & 1 \\
        \end{array}
    \right)
\end{equation*}
Analogously, the flow diagram at the level of pair transitions can be represented by an adjacency matrix $\mathscr{A}_P$. Remarkably, since the transition of a pair from a state to another is determined by the progress of an individual from a compartment to another, we can evaluate the matrix $\mathscr{A}_P$ from the matrix $\mathscr{A}_I$ as
\begin{equation*}
    \mathscr{A}_P = \mathscr{A}_I \otimes \mathscr{A}_I.
\end{equation*}
where the symbol $\otimes$ denotes the matrix direct product (also known as Kronecker product).
In components, matrix $\mathscr{A}_P$ is given as 
\begin{equation*}
    \left(\mathscr{A}_P\right)_{(i,\ell),(j,m)} = \left(\mathscr{A}_I\right)_{i,j}
    \left(\mathscr{A}_I\right)_{\ell,m}
\end{equation*}
which means that there is a link from pair $(i,\ell)$ to pair $(j,m)$ in the pairwise graph if and only if there exist both edges $(i,j)$ and $(\ell,m)$ in the individual graph.   
The set of all the possible (twenty-five) different pair states in the pairwise SEAIR model is 
shown in Fig.~\ref{fig:pairwise_transitions}, 
together with the flow diagram associated to the adjacency matrix $\mathscr{A}_P$. Again, as in 
Fig.~\ref{fig:SEAIR_model}, self-loops are not displayed. Finally, from this diagram, we can write the master equations for the pairwise SEAIR model as

\begin{widetext}
\begin{equation}
\begin{array}{lll}
    \langle SS \rangle_{t+1} &=& \langle SS \rangle_{t} - 2\Pi_{SS\rightarrow SE} - \Pi_{SS\rightarrow EE} \\[2pt]
    \langle SE \rangle_{t+1} &=& \langle SE \rangle_{t} + \Pi_{SS\rightarrow SE} - \Pi_{SE\rightarrow SA} - \Pi_{SE\rightarrow EA} - \Pi_{SE\rightarrow EE}  \\[2pt]
    \langle SA \rangle_{t+1} &=& \langle SA \rangle_{t} + \Pi_{SE\rightarrow SA} - \Pi_{SA\rightarrow EA} - \Pi_{SA \rightarrow ER} - \Pi_{SA \rightarrow SR} - \Pi_{SA \rightarrow EI} - \Pi_{SA \rightarrow SI} \\[2pt]
    \langle SI \rangle_{t+1} &=& \langle SI \rangle_{t} + \Pi_{SA\rightarrow SI} -  \Pi_{SI\rightarrow SR} - \Pi_{SI\rightarrow EI} - \Pi_{SI\rightarrow ER} \\[2pt]
    \langle SR \rangle_{t+1} &=& \langle SR \rangle_{t} + \Pi_{SA\rightarrow SR} + \Pi_{SI\rightarrow SR} - \Pi_{SR\rightarrow ER}\\[2pt]
    \langle EE \rangle_{t+1} &=& \langle EE \rangle_{t} + \Pi_{SS\rightarrow EE} + 2\Pi_{SE\rightarrow EE} - 2\Pi_{EE\rightarrow EA} - \Pi_{EE\rightarrow AA}\\[2pt]
    \langle EA \rangle_{t+1} &=& \langle EA \rangle_{t} + \Pi_{SE\rightarrow EA} + \Pi_{EE\rightarrow EA} + \Pi_{SA\rightarrow EA} - \Pi_{EA\rightarrow AA} - \Pi_{EA\rightarrow AI}
    - \Pi_{EA\rightarrow AR} - \Pi_{EA\rightarrow ER} - \Pi_{EA\rightarrow EI} \\[2pt]
    \langle EI \rangle_{t+1} &=& \langle EI \rangle_{t} +  \Pi_{SA\rightarrow EI}  + \Pi_{SI\rightarrow EI} + \Pi_{EA\rightarrow EI} - \Pi_{EI\rightarrow AI} - \Pi_{EI\rightarrow ER}  - \Pi_{EI\rightarrow AR}\\[2pt]
    \langle ER \rangle_{t+1} &=& \langle ER \rangle_{t} + \Pi_{SI\rightarrow ER} + \Pi_{EA\rightarrow ER} + \Pi_{SA\rightarrow ER} + \Pi_{SR\rightarrow ER} + \Pi_{EI\rightarrow ER} - \Pi_{ER\rightarrow AR}\\[2pt]
    \langle AA \rangle_{t+1} &=& \langle AA \rangle_{t} + 2\Pi_{EA\rightarrow AA} + \Pi_{EE\rightarrow AA} - \Pi_{AA\rightarrow II} - 2\Pi_{AA\rightarrow IR} - 2\Pi_{AA\rightarrow AI}  - 2\Pi_{AA\rightarrow AR} - \Pi_{AA\rightarrow RR} \\[2pt]
    \langle AI \rangle_{t+1} &=& \langle AI \rangle_{t} + \Pi_{AA\rightarrow AI} + \Pi_{EA\rightarrow AI} + \Pi_{EI\rightarrow AI} - \Pi_{AI\rightarrow IR} - \Pi_{AI\rightarrow RR}  - \Pi_{AI\rightarrow AR} - \Pi_{AI\rightarrow II} \\[2pt]
    \langle AR \rangle_{t+1} &=& \langle AR \rangle_{t} + \Pi_{AI\rightarrow AR} + \Pi_{ER\rightarrow AR} + \Pi_{EA\rightarrow AR} + \Pi_{AA\rightarrow AR} + \Pi_{EI\rightarrow AR}- \Pi_{AR\rightarrow IR} - \Pi_{AR\rightarrow RR} \\[2pt]
    \langle II \rangle_{t+1} &=& \langle II \rangle_{t} + \Pi_{AA\rightarrow II} + 2\Pi_{AI\rightarrow II} - 2\Pi_{II\rightarrow IR} - \Pi_{II\rightarrow RR}\\[2pt]
    \langle IR \rangle_{t+1} &=& \langle IR \rangle_{t} + \Pi_{II\rightarrow IR} + \Pi_{AA\rightarrow IR} +  \Pi_{AR\rightarrow IR} + \Pi_{AI\rightarrow IR} - \Pi_{IR\rightarrow RR}\\[2pt]
    \langle RR \rangle_{t+1} &=& \langle RR \rangle_{t} + \Pi_{II\rightarrow RR} + 2\Pi_{IR\rightarrow RR} + \Pi_{AA\rightarrow RR} + 2\Pi_{AI\rightarrow RR} + 2\Pi_{AR\rightarrow RR},
    \end{array}
    \label{eq:pair}
\end{equation}
\end{widetext}
where the term $\Pi_{XY \rightarrow X'Y'}$ represents the transition probability from the pair state $(X,Y)$ to the pair state $(X',Y')$. Note that, as $\langle XY \rangle_t = \langle YX \rangle_t$, we have that $\Pi_{XY \rightarrow X'Y'} = \Pi_{YX \rightarrow Y'X'}$. Due to the symmetries in the pair states, the equations for the states $(X',Y')$ and $(Y',X')$ are equivalent. For this reason,  we have only reported here the system of fifteen distinct equations (over the total of twenty-five) needed to fully characterize the system.

As for the individual-based model, we distinguish two classes of transition probabilities, i.e. linear and nonlinear ones. In particular, the class of nonlinear transition probabilities consists of all the terms involving at least one node in state $S$, as the probability of being infected (and consequently the probability of remaining susceptible) depends on the state of the nearest neighbors of the node. 

Again, let us first consider the linear transition probabilities. As an example we focus 
on the case of two symptomatic infected individuals, one of which recovers while the other remains infectious, i.e. on the transition from state $(I,I)$ to state $(I,R)$. The probability of such transition can be written as:
\begin{equation}
    \Pi_{II \rightarrow IR} = \langle II \rangle_t (1-\mu_I)\mu_I
    = \Pi_{II \rightarrow RI},
    \label{eq:linear_trans_pair}
\end{equation}

where the term $(1-\mu_I)\mu_I$ considers the two independent processes taking place in the pair transition, namely the recovery of the first node of the pair, occurring with probability $\mu_I$, and the persistence of the infection in 
the second, occurring with probability $(1-\mu_I)$. 

\begin{figure}[t]
    \centering
    \includegraphics[width=0.8\linewidth]{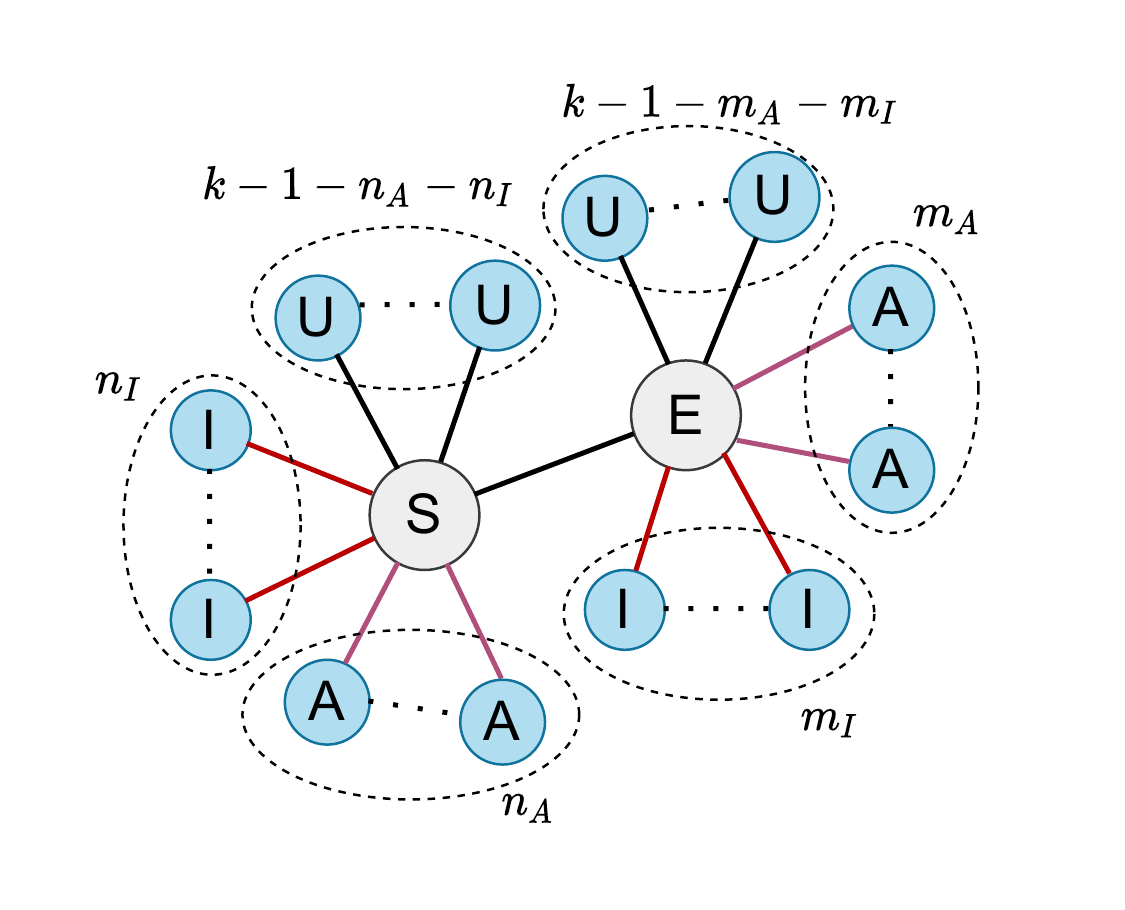}
    \caption{Graphical representation of the subgraph induced by a link $(i,j)$. Note that we are considering a configuration where each of the two nodes has $k$ neighbors and there are no triangular loops, so that $L=2k-2$. Here, 
node $i$ is in state $S$, node $j$ is in state $E$, 
    $n_A$ ($m_A$) indicates the number of neighbors of $i$ (of $j$) in state A, and, similarly, $n_I$ ($m_I$) those in state I.}
    \label{fig:subgraph_pairwise}
\end{figure}

As an example of nonlinear transition probability, let us consider the transition from state $(S,E)$ to state $(E,A)$, i.e. the case of a susceptible node connected to an exposed one, with the former that gets infected, while the latter becomes asymptomatic infectious. Similarly to the individual-based model formulation, as the infection of a susceptible node depends on the state of its neighbors, we first need to develop a node-level description of the process. Hence, let us denote the susceptible node as $i$, the exposed node as $j$, and derive the expression for the probability, $\Pi_{S_i E_j\rightarrow E_i A_j}$, that the pair $(i,j)$ moves from state $(S_i,E_j)$ to state $(E_i,A_j)$ at time $t$. We assume that the nodes $i$ and $j$ are connected on the contact graph $\mathcal{G}$, i.e. 
according to the adjacency matrix $G=\{g_{ij}\}$, 
and we consider the subgraph induced by the pair $(i,j)$. 
Such subgraph consists of nodes $i$ and $j$ and of all their 
$L$ neighbours. An example in which node $i$ is in state S, node $j$ is in state E and $k_i=k_j=k$ is shown in Fig.~\ref{fig:subgraph_pairwise} 

The transition probability can be written as:
\begin{equation}
    \begin{array}{ll}
    &\Pi_{S_i E_j \rightarrow E_i A_j} = \langle S_iE_jI_{h_1}U_{h_2}\ldots U_{h_L} \rangle_t [1-(1-g_{ih_1}\beta_I)]\\
    &+ \langle S_iE_jU_{h_1}I_{h_2}\ldots U_{h_L} \rangle_t [1-(1-g_{ih_2}\beta_I)]\alpha_{EA}\\
    & +\ldots\\
    &+ \langle S_iE_jU_{h_1}U_{h_2}\ldots I_{h_L} \rangle_t [1-(1-g_{ih_L}\beta_I)]\alpha_{EA}\\
    &+ \langle S_iE_jA_{h_1}U_{h_2}\ldots U_{h_L} \rangle_t [1-(1-g_{ih_1}\beta_A)]\alpha_{EA}\\
    &+\ldots\\
    &+ \langle S_iE_jU_{h_1}U_{h_2}\ldots A_{h_L} \rangle_t [1-(1-g_{ih_L}\beta_A)]\alpha_{EA}\\
    &+ \langle S_iE_jI_{h_1}I_{h_2}\ldots U_{h_L} \rangle_t [1-(1-g_{ih_1}\beta_I)(1-g_{ih_2}\beta_I)]\\
    &+\ldots\\
    &+ \langle S_iE_jI_{h_1}U_{h_2}\ldots I_{h_L} \rangle_t
    [1-(1-g_{ih_1}\beta_I)(1-g_{ih_L}\beta_I)]\\
    &+ \langle S_iE_jI_{h_1}A_{h_2}\ldots U_{h_L} \rangle_t [1-(1-g_{ih_1}\beta_I)(1-g_{ih_2}\beta_A)]\\
    &+\ldots\\
    &+ \langle S_iE_jA_{h_1}A_{h_2}\ldots A_{h_L} \rangle_t [1-\prod_{n=1}^{L}(1-g_{ih_n}\beta_A)]\alpha_{EA},
    \end{array}
    \label{eq:transition_pair}
\end{equation}

\noindent where $h_1$, $h_2$, ..., $h_L$ label the $L$ nodes in the neighborhood of link $(i,j)$.   

Each term on the right-hand side of  Eq.~(\ref{eq:transition_pair}) is given by  
the probability that the subgraph induced by $(i,j)$ is in a certain configuration, conditional to $(i,j)$ being in state $(S_i,E_j)$, times a term which, in turn, is given by the product of the probability that node $i$ gets infected and the probability that node $j$ becomes infectious over the next time step. The different terms in Eq.~(\ref{eq:transition_pair}) 
take into account all the possible different states of the neighbourhood of link $(i,j)$.

Similarly to what observed in the individual-based model, given the current form of the transition probability, also Eqs.~(\ref{eq:pair}) are exact but not closed. This time, however, we close the system equations at the level of links by assuming statistical independence in the states of node pairs. In other words, to close the system we need to find an approximating function $F$ that allows to write the higher-order joint probability as 
\begin{equation}
\begin{array}{l}
\label{eq:SE_node_closure}
\langle S_iE_j X_{h_1}\ldots Y_{h_{L-1}} Z_{h_L} \rangle_t \approx
F(\langle S_{i} E_{j} \rangle_t,\ldots,\langle Y_{h_{L-1}} Z_{h_L} \rangle_t)
\end{array}
\end{equation}
At variance with the individual-based model, for which 
the expression of $F$ is straightforward, several moment closures exist for the pairwise model, whose quality depends on the topology of the underlying contact network \cite{pellis2015exact}. Here we consider a closure which is exact under the assumption that the network 
contains no cycles of any order.
Though this assumption does not hold in real finite systems, the closure above provides a valuable approximation for the analysis of dynamic processes in large sparse networks
(such as sparse random regular graphs and Erd\"os-Renyi random graphs), in which the number of cycles is negligibly small.
Taking into account these considerations, the subgraph $\mathcal{G}_{ij} = (\mathcal{V}_{ij},\mathcal{E}_{ij})$ induced by $(i,j)$ is in the form shown in   Fig.~\ref{fig:subgraph_pairwise},
and, according to Ref.~\cite{frasca2016discrete}, the following approximating function $F$ can be adopted
\begin{equation}
\begin{array}{l}
F(\langle S_i E_j \rangle_t,\ldots,\langle Y_{h_{L-1}} Z_{h_L} \rangle_t) = \frac{\prod_{(n,m)\in \mathcal{E}_{ij}}\langle N_n M_m\rangle_t}{\langle S_i\rangle_t^{k_i-1}\langle E_j\rangle_t^{k_j-1}},
\end{array}
\label{eq:node_level_closure}
\end{equation}
Note that the numerator consists in the product of the state probabilities of each link in $\mathcal{E}_{ij}$. Note also that $\langle S_i\rangle_t$ and $\langle E_j\rangle_t$ are the marginals probabilities evaluated from Eq.~(\ref{eq:marginal}) and that the closure in Eq.~(\ref{eq:node_level_closure}) is consistent with e marginal probabilities it is constructed from, namely   
\begin{equation}
\begin{array}{l}
\sum\limits_{X_{h_1}, Y_{h_2}, \dots Z_{h_L}\in\Omega} 
F(\langle S_i E_j \rangle_t,\ldots,\langle Y_{h_{L-1}} Z_{h_L} \rangle_t)=
\langle S_i, E_j \rangle_t
\end{array}
\end{equation}
The joint probabilities in Eq.~(\ref{eq:transition_pair}) can then be rewritten using the expression of the approximating function in Eq.~(\ref{eq:node_level_closure}). We can finally introduce the homogeneous mixing hypothesis to formulate the master equations at the level of population, which permits to drop the node indices in the probability terms. Assuming each node to be connected to $k$ neighbors, after some manipulation=s (detailed in Appendix B), we can write the transition probability $\Pi_{SE\rightarrow EA}$ as
\begin{equation}
    \begin{array}{l}
    \Pi_{SE\rightarrow EA} \approx \langle SE \rangle_t \sum\limits_{p=1}^{k-1}\sum\limits_{n=0}^{p} \frac{\langle SA \rangle_t^{n}\langle SI \rangle_t^{p-n}\langle SU \rangle_t^{k-1-p}}{\langle S\rangle_t^{k-1}}\\
    \times{k-1 \choose p}{p \choose n}[1-(1-\beta_A)^n(1-\beta_I)^{p-n}]\\ \times\sum\limits_{q=0}^{k-1}\sum\limits_{m=0}^{q}\frac{\langle EA \rangle_t^{m}\langle EI \rangle_t^{q-m}\langle EU \rangle_t^{k-1-q}}{\langle E\rangle_t^{k-1}}{k-1 \choose q}{q \choose m}\alpha_{EA}.
    \end{array}
    \label{eq:double_sum}
\end{equation}
This can be further simplified, so to write an expression which is similar to the one obtained for the individual-based model, i.e.\ Eq.~(\ref{eq:PiSEchiusaInd}). We have
\begin{equation}
\begin{array}{l}
    \Pi_{SE\rightarrow EA} \approx \langle SE \rangle_t \left[1-\left(1-\beta_A\frac{\langle SA \rangle_t}{\langle S \rangle_t} - \beta_I\frac{\langle SI \rangle_t}{\langle S \rangle_t}\right)^{k-1}\right]\\
    \times \alpha_{EA}.
\end{array}
\label{eq:pairwise_population_SE_EA}
\end{equation}
Note that as the transition of an individual from state E to state A is independent on the state of the neighbors, the transition probability does not depend on the state probabilities $\langle EX\rangle$ but on $\alpha_{EA}$ only.

The other transition terms appearing in Eq.~(\ref{eq:pair}) are in the form of Eq.~(\ref{eq:linear_trans_pair}), if linear, or of Eq.~(\ref{eq:pairwise_population_SE_EA}), if nonlinear, and are derived following similar arguments. They are fully reported in Appendix C. Eq.~(\ref{eq:pair}) along with Eq.~(\ref{eq:linear_trans_pair}) and Eqs.~(C1)-(C10) represent a closed set of equations for the pairwise population-level SEAIR model.

We conclude the Section, as done for the individual-based model, here we report the expression of the basic reproduction number for the pairwise SEAIR model, evaluated through the NGM approach (see Appendix D). In this case we have
\begin{equation}
\label{eq:R0_pair}
    \mathcal{R}_0 = \frac{\left(k - 1\right) \{\alpha_{AI} \beta_{I} \left(1 - \beta_{A}\right) + \beta_{A} \left[1- \left(1 - \beta_{I}\right) \left(1 - \mu_{I}\right)\right]\}}{\left[1- \left(1 - \beta_{A}\right) \left(1 - \alpha_{AI} - \mu_{A}\right) \right] \left[1- \left(1 - \beta_{I}\right) \left(1 - \mu_{I}\right) \right]}
\end{equation}

\section{Reproducing the evolution of an epidemic}
\label{section:epidemic_evolution}

\begin{figure}[t!]
    \centering
    \includegraphics[width=\linewidth]{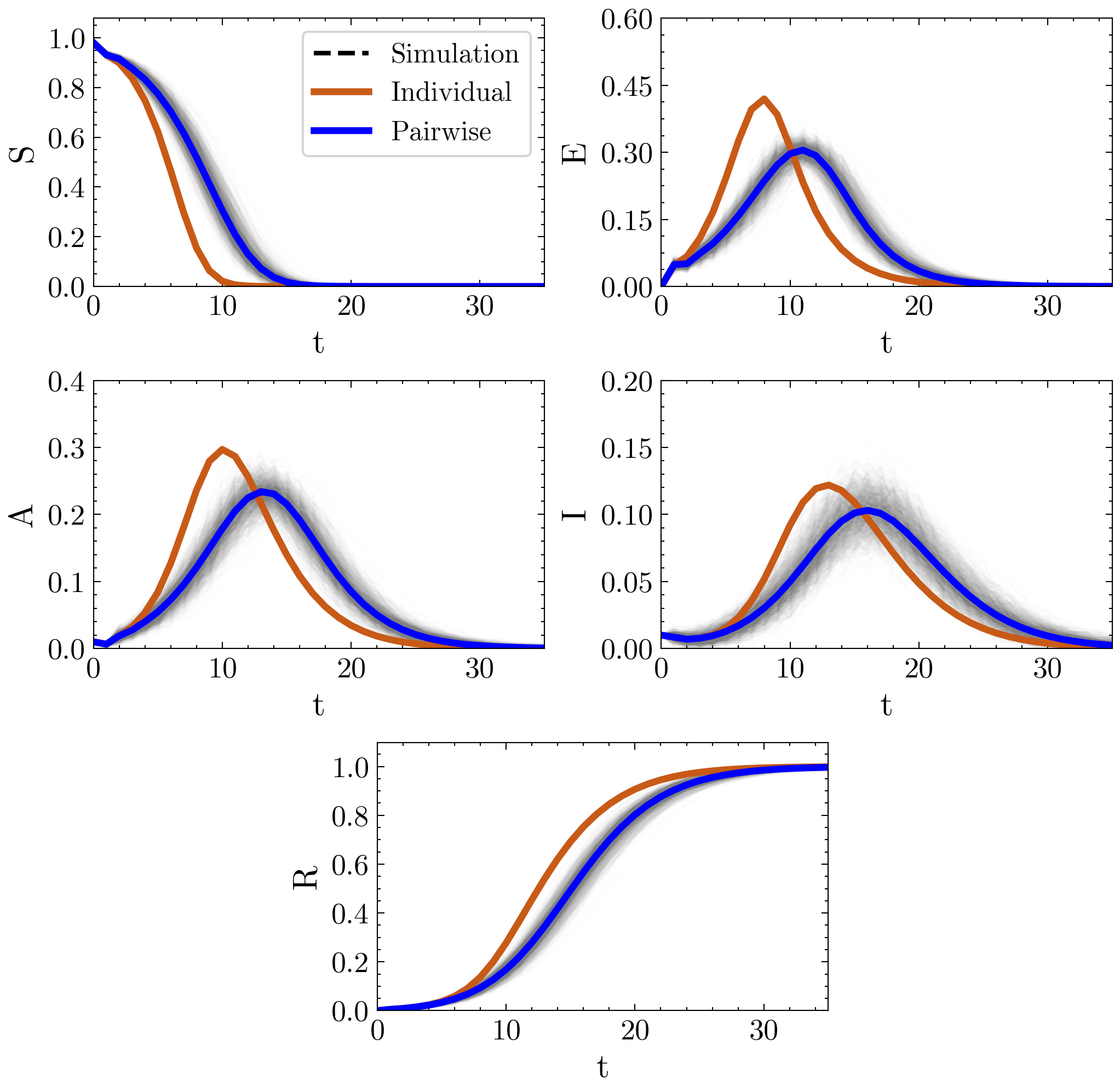}
    \caption{Comparison between stochastic simulations, individual-based model (orange solid line) and pair-based model (blue solid line) for a random $r$-regular graph with $N=500$ nodes and $r=5$. Both the numerical simulations and the models dynamics are evaluated with $\beta_I=0.4$, $\beta_A=0.6$, $\alpha_{EA} = 0.3$, $\alpha_{AI} = 0.2$, $\mu_A = 0.15$, $\mu_I = 0.3$, $S(0) = 99.98$, $E(0) = 0$, $A(0) = 0.01$, $I(0) = 0.01$, $R(0) = 0$. The stochastic simulations are represented as solid gray lines, while the average is plotted as a dashed black line.}
    \label{fig:reproducing_random_regular}
\end{figure}

The individual-based and the pairwise model provide an approximate description of a disease spreading over a network. Our goal is
now to study to which extent, and under which conditions, these approximate descriptions can capture the time evolution of an epidemic outbreak.  
%
%
We will do this by performing different types of test. 
In this section we will first concentrate on the simpler case in which the parameters that regulate the disease spreading are known, while in the next section we will consider the case in which such parameters need to be extracted from the data. 

Under the assumption that the model parameters are given, we follow the time course of the disease using the different models, starting from the same initial conditions, and compare their capacity in predicting different aspects of a disease outbreak, such as the time of the epidemic peak, its height, and the final number of infected individuals. 
To illustrate the results of our analysis, we focus on two different network topologies, namely a random $r$-regular graph and an Erd\"os-Renyi (ER) graph \cite{latora2017complex}. We start with random regular graphs that, for their characteristics, are the network topologies that best match the hypothesis under the derivation of the closure (\ref{eq:node_level_closure}). Indeed, in a random $r$-regular graph each node has the same degree, equal to $r$, and the degree distribution is a Kronecker delta, i.e., $P(k) = \delta_{r,k}$. In addition, the expected number of triangles is asymptotically equal to $(r-1)^3 / 6$ \cite{bollobas2001random}, therefore, their effect is negligible in large networks.

\begin{figure}[t!]
    \centering
    \includegraphics[width=\linewidth]{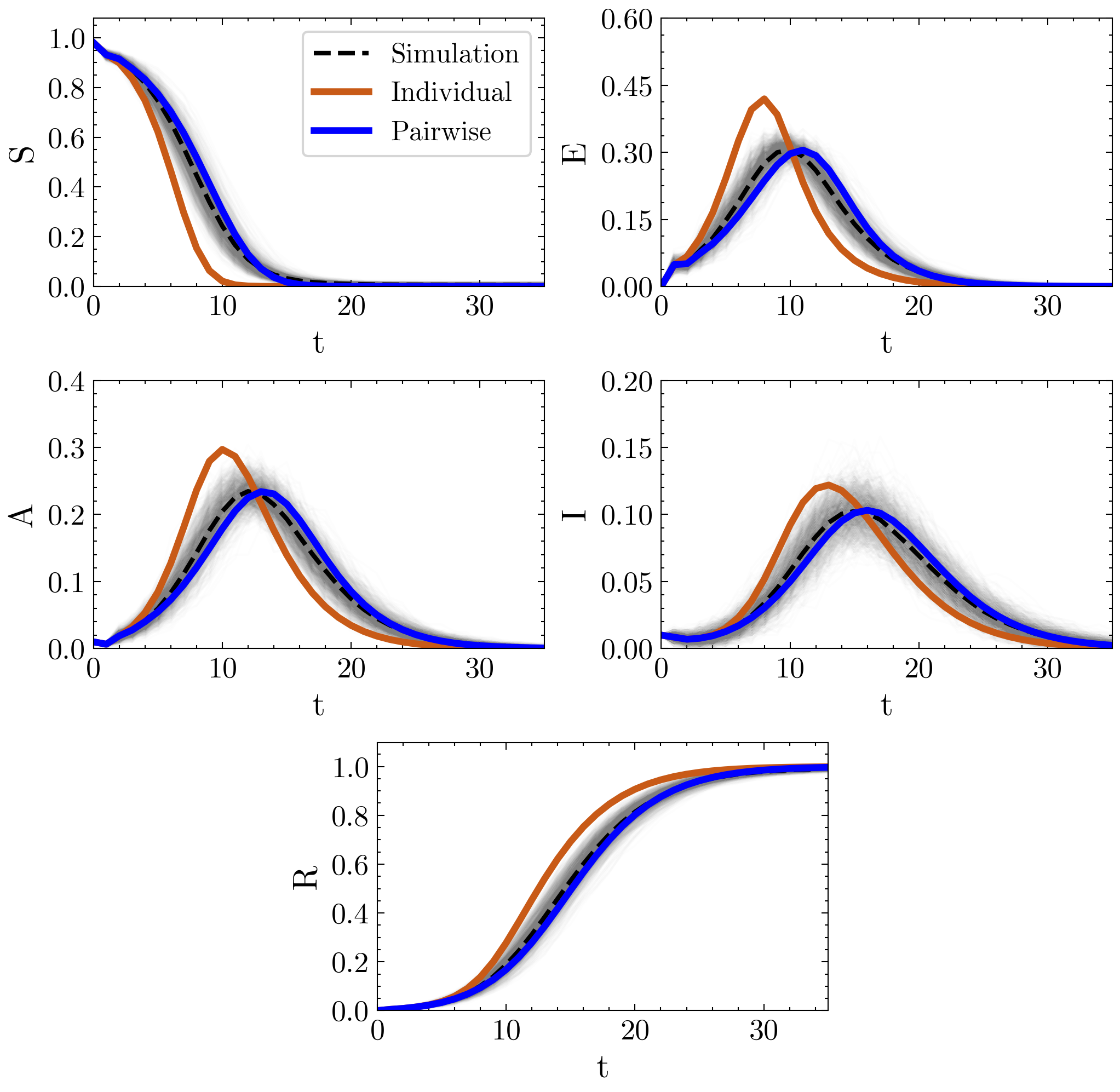}
    \caption{Comparison between stochastic simulations, individual-based model (orange solid line) and pair-based model (blue solid line) for an Erd\"os-Renyi graph with $N=500$ nodes and $p=0.01$. The numerical simulations and the model dynamics are evaluated using the same settings of Fig.~\ref{fig:reproducing_random_regular}. The stochastic simulations are represented as solid gray lines, while the average is plotted as a dashed black line.}
    \label{fig:reproducing_ER}
\end{figure}

To simulate the epidemic process on a network, we consider, for each time step, the kinetic equations at the level of single nodes. At each iteration we inspect all the nodes in states E, A and I. For each of them, we run a Bernoulli process to determine whether the node transits to another state or not. Additionally, for each of the nodes in states A or I we consider each of their susceptible neighbors to determine if it gets infected or not, with probability $\beta_A$ or $\beta_I$, respectively. These \emph{stochastic simulations} are performed on networks with $N=500$ nodes and $r=5$. The initialization of the stochastic process is done by uniformly sampling $2\%$ of the nodes and assuming them to be infected. Half of these infectious nodes were initialized as asymptomatic infectious  individuals, while the other half as  symptomatic ones. All the remaining nodes were set in the susceptible state. For each case study investigated, a number $M$ of runs of the stochastic simulations is considered, each time re-initializing the network by randomly choosing the nodes to set in the A or I state.
\begin{figure*}[t!] 
    \centering
    \includegraphics[width=\linewidth]{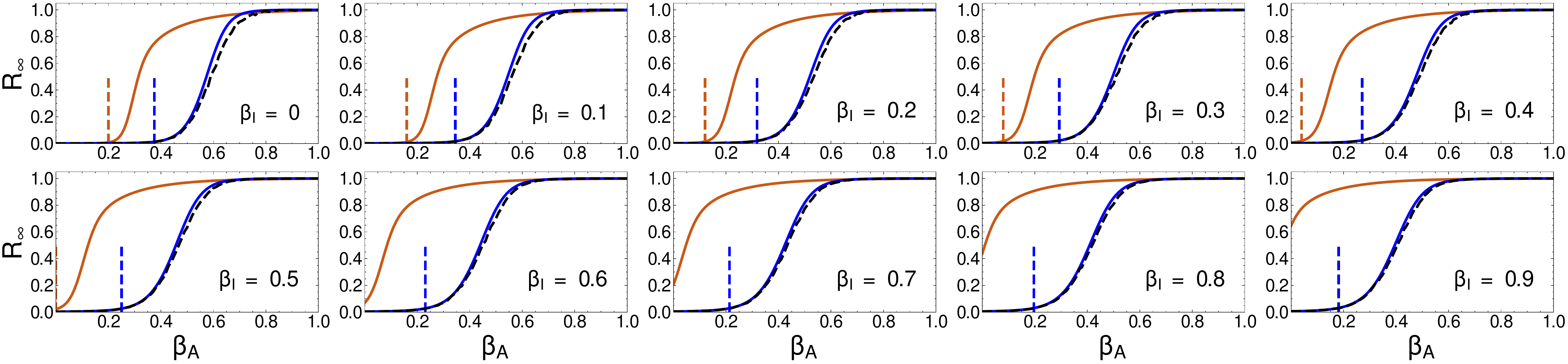}

    \caption{Final size of the recovered population $R_\infty$ as a function of $\beta_A$, for different of $\beta_I$. The numerical simulations are performed on a random $r$-regular graphs with $N = 2000$  nodes and $r = 3$, with $\alpha_{EA} = 0.3$, $\alpha_{AI} = 0.2$, $\mu_A = 0.4$, $\mu_I = 0.5$. 
    The solid lines represent the value of $R_\infty$ as predicted by the individual-based (orange) and by the pairwise model (blue), while the black dashed line represents the average value of $R_\infty$ over $M = 1000$ stochastic simulations. The vertical dashed lines represent the value of $\beta_A$ at which the value of $\mathcal{R}_0$ is equal to one for both the individual-based (orange) and pairwise model (blue).}
    \label{fig:r_eq_10_panels}
\end{figure*}

As an example, the evolution of the disease obtained for a given set of parameters, namely $\beta_I=0.4$, $\beta_A=0.6$, $\alpha_{EA} = 0.3$, $\alpha_{AI} = 0.2$, $\mu_A = 0.15$, $\mu_I = 0.3$, $S(0) = 99.98$, $E(0) = 0$, $A(0) = 0.01$, $I(0) = 0.01$, $R(0) = 0$, is shown in Fig.~\ref{fig:reproducing_random_regular}. Each of the five dynamical variables of the SEAIR model, namely the percentage of nodes respectively in the S, E, A, I and R state, are reported. The results of $M=1000$ stochastic simulations are represented as grey lines, while the average is plotted as a dashed black line. The evolution predicted by the individual- and the pair-based model are also reported as orange and blue solid lines, respectively, with the blue solid line, which appears to be almost superimposed to the dashed black line. 
We observe that the individual-based model substantially overestimates the number of  infections, whereas the dynamics predicted by the pairwise model well reproduces the average behavior of the stochastic simulations. These findings are in agreement with the results obtained for the SIR model in Ref.~\cite{frasca2016discrete}, supporting the conclusion that pairwise models are very good approximations of  the dynamics of an epidemic on random regular graphs independently from the specific type of epidemic process considered.

In order to show how the two models perform on a network that deviates from the assumptions underlying the derivation of the pairwise model, we here consider ER random graphs. In this case, the degree distribution is notably binomial peaked at $\langle k \rangle = Np$, where $p$ is the probability that a link between two nodes exists. Notice that, while for random $r$-regular graphs the degree distribution is a Kronecker delta, for ER graphs the variance on the node degrees is nonzero, i.e., $\sigma^2 = Np(1-p)$. Similarly to the case of random regular graphs, triangles can also be neglected in ER random graphs, as their expected number is asymptotically equal to $\langle k \rangle^3 / 6$ \cite{latora2017complex}.
Again, we fix the set of parameters (e.g.\ the same as in the previous example) and we compare the time evolution of the epidemics obtained for the individual-based and pairwise models to the results of $M=1000$ runs of the stochastic process on  
ER random networks with $N=500$ nodes and $p=0.01$. Fig.~\ref{fig:reproducing_ER} shows that, despite being less accurate than in the previous example, the dynamics of the pairwise model is still in good agreement with the average behavior of the stochastic simulations, whereas the individual-based model fails to reproduce several features of the time evolution of the epidemic outbreak. Ideed for both network topologies considered above, the comparison between the models and the stochastic simulations shows that the individual-based model overestimates the total number of carriers of the disease, and underestimates the peak time and the duration of the epidemic process. This is consistent with the results of a more systematic analysis that we have carried out by varying the model parameters. 

\begin{figure}[t!] 
    \centering
    \includegraphics[width=\columnwidth]{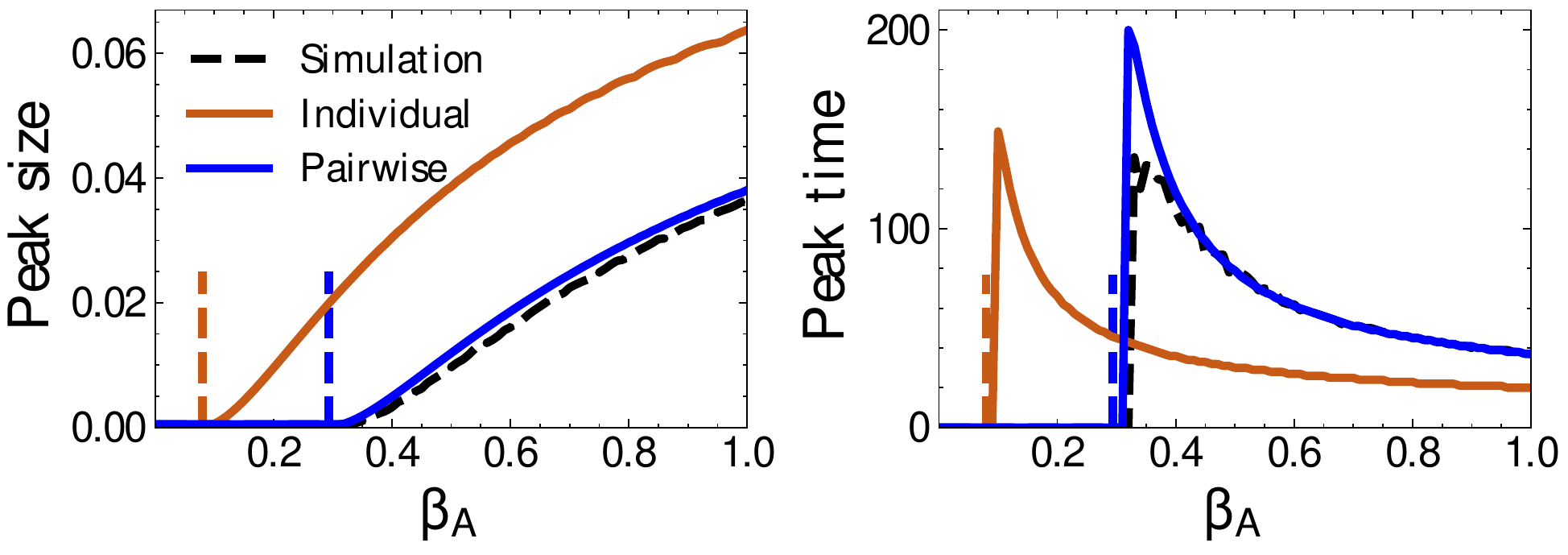}
    \caption{Size (left) and time (right) of the peak in the number of symptomatic individuals, as a function of $\beta_A$, for $\beta_I = 0.3$. The numerical simulations are performed with the same settings of Fig.~\ref{fig:r_eq_10_panels}. The solid colored lines refers to the individual-based (orange) and the pairwise model (blue), while the black dashed line represents the stochastic simulations. The vertical dashed lines represent the value of $\beta_A$ at which $\mathcal{R}_0=1$ for both the individual-based (orange) and pairwise model (blue).}
    \label{fig:size_and_time}
\end{figure}

Figs.~\ref{fig:r_eq_10_panels} and \ref{fig:size_and_time} illustrate some of the results of this analysis for the case of random $r$-regular graphs, where, in particular, we have varied the infection probability for the asymptomatic individuals in the range $\beta_A \in [0,1]$ for different values of $\beta_I$, and monitored several macroscopic quantities of interest, such as the final size of the recovered population, and the value and time of the peak in the number of symptomatic infected individuals. Note that, in this case, we have considered a larger network, i.e.\ $N=2000$, and a smaller fraction of initial infected individuals, i.e.\ $10^{-3}$, so to better analyze the behavior of the system at an early stage of the epidemic outbreak.

Let us first study the behavior of the final size of recovered population, i.e. $R_\infty$ (Fig.~\ref{fig:r_eq_10_panels}). Coherently with the previous results, as the individual-based model overestimates the number of infections occurring at each time step, the final size of recovered individuals is close to one even for relatively small values of $\beta_A$ (see for instance the case in which $\beta_I=0.9$), failing to predict the exact value $R_\infty$. At the same time, the pairwise model provides a good prediction of $R_\infty$ for different values of $\beta_I$ and in the entire range of $\beta_A$ considered. 
%
Similarly, concerning the size and the time of the peak of symptomatic individuals, (Fig.~\ref{fig:size_and_time}), we observe that pairwise model is able to give better predictions compared to the individual-based one. Again, the individual-based model overestimates the size of the peak and anticipates its actual time, while the pairwise model is in good agreement with the numerical simulations. The differences between the peak time observed in the stochastic simulations and that predicted by the pairwise model around $\beta_A \approx 0.32$ are mainly due the phenomenon of stochastic fade-out \cite{britton2015five} that is more pronounced for $\mathcal{R}_0$ slightly greater than one.

A further remarkable aspect of the pairwise model is that it provides a more precise estimation of the critical values of $\beta_A$ and $\beta_I$ at which the epidemic outbreak occurs. In both figures and for both deterministic models, we also display the value of $\beta_A$ for which, given the value of the other parameters, the basic reproduction number $\mathcal{R}_0$ goes to zero. We observe that the pairwise model seems to correctly predict the threshold value of $\beta_A$, for different $\beta_I$. Contrarily, the individual-based model predicts smaller threshold values, meaning that there is a range of $\beta_A$ for which the model wrongly forecasts the onset of an epidemic outbreak. In particular, as shown in Fig.~\ref{fig:r_eq_10_panels}, for $\beta_I>0.4$ the individual-based model does not admit a stable disease-free equilibrium, i.e. the infection always spreads for any value of $\beta_A$, while the pairwise model correctly predicts the epidemic threshold.

Altogether, the results presented in this section clearly show that taking into account the correlations in the network of contacts is a fundamental step for an accurate description of the spreading of an epidemic in a population. Indeed 
the pairwise model outperforms the individual-based one, especially in reproducing the real evolution of the disease. Still, it is worth noting that the analysis presented above has been carried out under the ideal condition that all parameters and variables are known. In practice,
however, this can be an oversimplifying and unrealistic hypothesis. For instance, some epidemiological parameters, such as the infection probabilities, may be hard to measure directly, and thus need to be estimated from empirical data on the disease spreading. Furthermore, some of the system variables can be unmeasurable. For example, it may not be possible to trace infected individuals during the incubation period, as the carriers themselves may be unaware of having been exposed to the disease. 
These considerations motivate the analysis that  will be presented in the next section, where we will compare the deterministic models adopting a different perspective. Instead of assessing the discrepancies in the dynamics given the epidemiological parameters, we will fit the models to data from numerical simulations and analyze the differences in the predictions of both the parameters of the model and the evolution of the state variables.  





\section{Dealing with incomplete information \label{section:fitting}}

In this section we concentrate on the case in which the epidemiological parameters are not known 
{\em a priori}, but need to be extracted from data. We will consider three separate cases in increasing order of complexity. 
We first start with the hypothesis that all the variables are known and they are so in the whole time interval considered. Then, we examine the case in which all the variables are known, but only up to a certain time. Finally, we assume that only a subset of the system variables is effectively available for fitting. For each of the three cases, we consider the problem of determining the parameters by fitting the available data and, whenever appropriate, that of predicting the temporal evolution of the system variables. The main focus would be again to compare the results of the individual-based and of the pairwise model.

\subsection{Identifying the epidemiological parameters}

We first study the problem of evaluating the epidemiological parameters when all the state variables are known for the entire course of the outbreak. As epidemic spreading models are often applied to the study of novel infectious diseases, they can be crucial for the determination of the more elusive epidemiological parameters, such as the infection probabilities or the incubation period, for which direct measurements can be difficult to perform. For this reason it is fundamental to assess the reliability of the estimation of the epidemiological parameters when using a model to fit the data. Here, we perform a series of numerical experiments aimed at comparing the individual-based and the pairwise deterministic SEAIR model in performing this task. We proceed in the following way. We fix a set of epidemiological parameters $\mathbf{p} = (\beta_A,\beta_I,\alpha_{EA},\alpha_{AI},\mu_A,\mu_I)$, which are assumed to be unknown to the deterministic models and need to be determined through the fit (note that, instead, the initial conditions are assumed to be known quantities). The synthetic data are obtained by running a series of microscopic simulations on a given contact network and averaging the results over $M$ different realizations of the stochastic process. The time-series obtained in this way plays the role of the empirical data that the deterministic models have to reproduce at their best. 
Note that, since real data usually consists in the daily/weekly number of individuals in a given state \cite{pcdata}, as synthetic time-series we consider the temporal evolution of the fractions of \emph{individuals} in each compartment, which we will denote with $\overline{X}_t$. Therefore, to determine the epidemiological parameters, we have to fit the corresponding quantities $\langle X \rangle_t$ to the synthetic data $\overline{X}_t$. It is important to remark that for the individual-based model these correspond to the state variables of the system, whereas for the pairwise model they are function of the state variables (the pair probabilities), through relation \eqref{eq:marginal}.
For each of the two models we perform an optimization procedure to derive the set of parameters $\hat{\mathbf{p}}$ that yield the smallest fitting error. The model parameters are estimated by minimizing the root mean squared error (RMSE)
\begin{equation}
    \mathcal{E}_{\rm fit} = \sqrt{\frac{1}{|\Omega'|T}\sum_{X\in\Omega'} \sum\limits_{t=1}^{T}( \langle X\rangle_t - \overline{X}_t )^2}
\end{equation}  
between the time-series produced by the deterministic model, generically represented as $\langle X \rangle_t$, and the corresponding average $\overline{X}_t$ of the microscopic simulations. We denote with $\Omega'$ the set of compartments to which we fit the models, which in general can differ from $\Omega$. In this first example, we will consider $\Omega' = \Omega$, so that $|\Omega'|=5$. 
The length of the simulation, $T$, is chosen such that the system has reached a stationary state, i.e. the epidemic outbreak has ended, as every infected individual has eventually recovered. 
Finally, as a measure of the discrepancy between the model estimation and the actual parameters we evaluate the quantity $\mathcal{D}(\mathbf{p}) = \|\mathbf{p} - \hat{\mathbf{p}}\|/\sqrt{6}$, where $\| \mathbf{v} \|$ denotes the Euclidean norm and 6 is the number of components in vector $\mathbf{p}$ \footnote{Note that all parameters are of the same order of magnitude and homogeneous, thus there is no need to use a weighted average in the computation of the error.}. 

\begin{figure}[t]
    \centering
    \includegraphics[width=\linewidth]{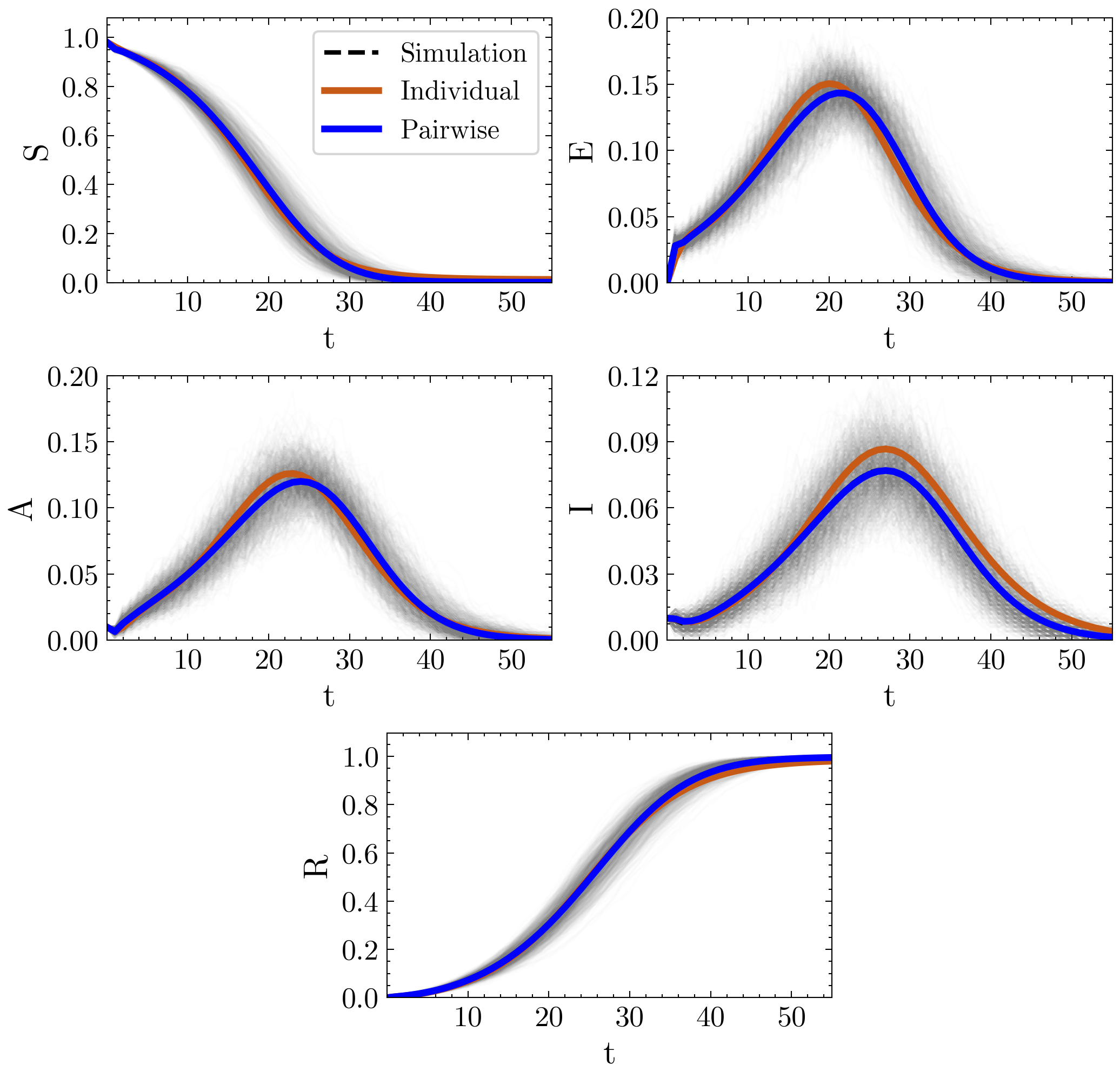}
    \caption{Comparison between stochastic simulations, individual-based model (orange solid line) and pair-based model (blue solid line) when all the state variables are fitted over the entire course of the epidemic. The stochastic simulations are run on a random $r$-regular graph, with the same settings of Fig.~\ref{fig:reproducing_random_regular}. The stochastic simulations are represented as solid gray lines, while the average is plotted as a dashed black line.
    }
    \label{fig:fit_5a}
\end{figure}

\begin{figure*}[t]
    \centering
    \includegraphics[width=\linewidth]{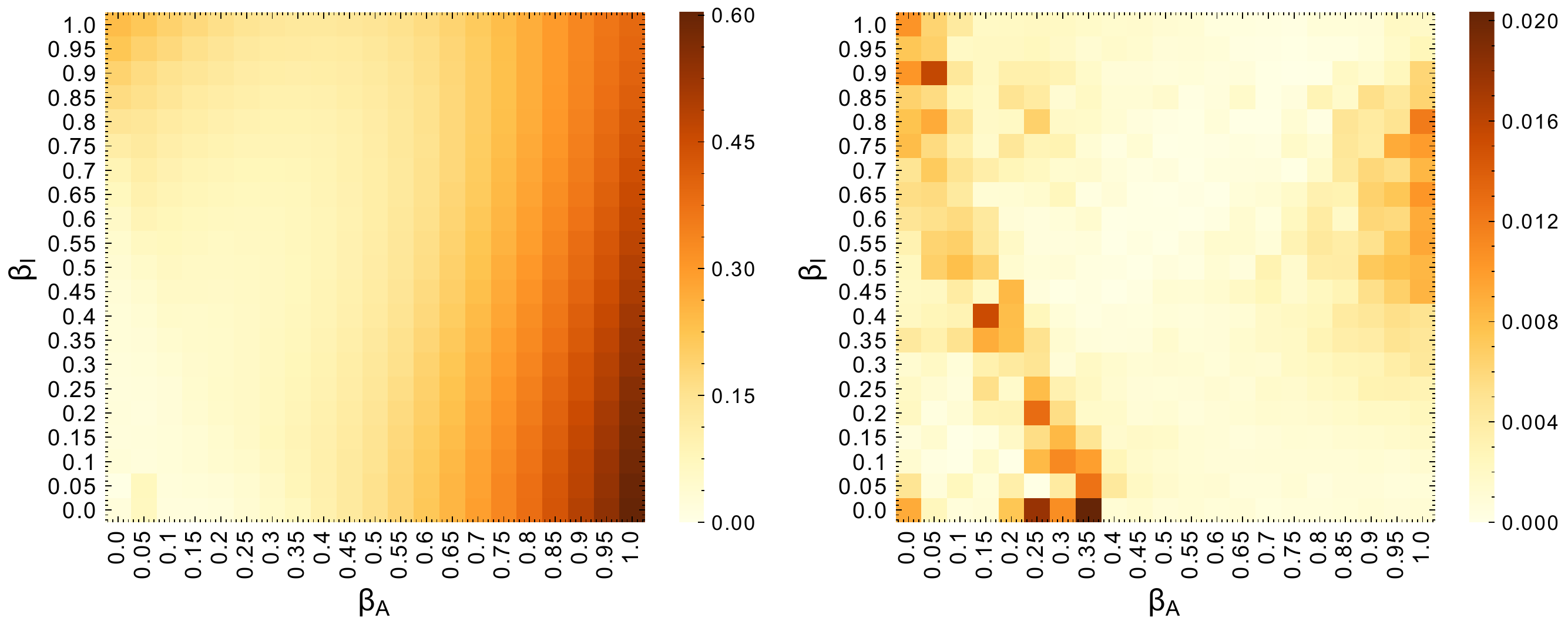}
    \caption{Error $\mathcal{D}(\mathbf{p})$ in the estimation of the epidemiological parameters as a function of $\beta_{A}$ and $\beta_{I}$ for the individual-based (left) and the pairwise (right) SEAIR models. The deterministic models are fitted to stochastic simulations performed on a $r$-regular graphs with $N=500$ and $r=3$, with $\alpha_{EA} = 0.3$, $\alpha_{AI} = 0.2$, $\mu_A = 0.15$ and $\mu_I = 0.3$.}
    \label{fig:heatmap_all_parameters}
\end{figure*}

An example of the results is shown in Fig.~\ref{fig:fit_5a}, where the model predictions are reported together with the numerical simulations in the case of random $r$-regular graphs with $N=500$ nodes and $r=3$ and of the set of parameters listed in the figure caption.  
Averages over $M=1000$ runs of the stochastic microscopic simulation have been considered. 
The system is initialized by picking a number of randomly selected nodes corresponding to the 2\% of the total population and setting their initial state, either as asymptomatic (A) or infected (I) with equal probability. 
\label{section:epidemic_parameters}
Fig.~\ref{fig:fit_5a} shows that the dynamical evolution of the state variables generated by both the individual-based and the pairwise model are in good agreement with the stochastic simulation, with the pairwise model still providing better results. Indeed, for the individual-based model we obtain $\mathcal{E}_{\rm fit}^{\rm ind}=9.0\cdot 10^{-5}$, while for the pairwise model we have $\mathcal{E}_{\rm fit}^{\rm pair}=3.7\cdot 10^{-7}$.
In addition, the parameters extracted with the individual-based model significantly deviate from the ``real ones'' used to generate the data to fit.
Conversely the trajectories of the pairwise model that best fit the data are obtained with parameters close to those used in the microscopic simulation. 
If we consider, for instance, the infection probabilities, which have been set to $\beta_A=0.6$ and $\beta_I=0.4$, we find that the parameters evaluated through the pairwise model ($\hat{\beta}_A=0.59$, $\hat{\beta}_I=0.35$) are in a good agreement with the epidemiological ones, whereas the estimates of the individual-based model ($\hat{\beta}_A=4.32 \cdot 10^{-6}$, $\hat{\beta}_I=0.63$) substantially differ from them. Now, if we consider the value of $\mathcal{D}(\mathbf{p})$, for the pairwise model we find $\mathcal{D}_{\rm pair}(\mathbf{p})=9.3\cdot10^{-4}$, whereas for the individual-based model we obtain $\mathcal{D}_{\rm ind}(\mathbf{p})=0.17$.
Hence, both the fitting error $\mathcal{E}_{\rm fit}$ and the discrepancy in the estimation of the parameters $\mathcal{D}(\mathbf{p})$ obtained by adopting the pairwise model are about two orders of magnitude smaller than the corresponding ones deduced by the individual-based model, so we conclude that the pairwise approximation provides a more reliable ``prediction'' of the epidemiological parameters.

These results are consistent with the conclusions of the previous section. In fact, when the deterministic models are informed of the epidemiological parameters, then the dynamics of the pairwise model closely matches the microscopic simulations, whereas the individual-based model largely overestimates the number of infections occurring at each time step. In the case considered in this section in which the epidemiological parameters are assumed to be unknown, both models are able to generate time-series that are close to the real time evolution of the epidemics. But in order to do so, the individual-based model has to use a set of parameters that significantly differs from that of the stochastic simulations. Still, the individual-based model has a higher fitting error. In the pairwise case, instead, when the parameters are known, the model behavior is close to that of the microscopic simulations, such that, when the parameters are obtained through fitting, their estimates slightly differ from those used to generate the data.

Fig.~\ref{fig:fit_5a} provides an illustrative example obtained for a single set of the epidemiological parameters. 
As the error in the parameter estimation does depend, in general, on the vector $\mathbf{p}$ itself, we have repeated the previous analysis on the same contact network represented by a random $r$-regular graph with $N=500$ and $r=3$, but considering different sets of epidemiological parameters. 
In particular, we have systematically varied two parameters of the system, $\beta_A$ and $\beta_I$, which are the ones ruling the disease transmission. For each pair $(\beta_A, \beta_I)$ we have calculated the average over $M=1000$ runs of the stochastic epidemic process, applied the fitting procedure for the two deterministic models, and computed the error $\mathcal{D}(\mathbf{p})$.

Fig.~\ref{fig:heatmap_all_parameters} displays $\mathcal{D}(\mathbf{p})$ as a function of $\beta_A$ and $\beta_I$. Two things are worth noticing. Firstly, the estimation error for the pairwise model (on the right) is generally lower compared to that of the individual-based model (on the left). Second, the value of $\mathcal{D}(\mathbf{p})$ for the individual-based model depends smoothly on $\mathbf{p}$, whereas, for the pairwise model, such a dependence appears to be strongly affected by random fluctuations, because of its much smaller value.

Overall, the previous results show that the pairwise approximation provides a more reliable estimation in the entire space of parameters. This suggests that the use of a pairwise model is preferable even when the epidemiological parameters of an  infectious diseases are not known. 

\subsection{Forecasting the epidemic evolution}
\label{section:predictive_capability}
So far, we have seen how to extract the epidemiological parameters when data on the entire course of the epidemic spreading are available. 
%
However, mathematical models 
%
are also useful to forecast the evolution of a  disease spreading 
\cite{estrada2020covid,kucharski2020calculating,vespignani2020modelling}. For instance, they play a crucial role in policy making, as 
their predictive power allows to estimate in advance the possible effects of different containment measures.
On the grounds of this, here we compare the capability of the individual-based and of the pairwise models in forecasting the progress of an epidemic when empirical data is only available for a limited time interval. 
Similarly to the previous case, we assume all the epidemiological parameters $\mathbf{p} = (\beta_A,\beta_I,\alpha_{EA},\alpha_{AI},\mu_A,\mu_I)$ are unknown quantities to be determined by fitting the deterministic models to the numerical simulations. This time, however, instead of considering the time-series $\overline{S}_t$, $\overline{E}_t$, $\overline{A}_t$, $\overline{I}_t$, and $\overline{R}_t$, over the entire time range $T = 55$, we fit the models over a limited time interval $[0,T_{\rm fit}]$. Then, we compare how the SEAIR models predict the evolution of the state variables in the remaining time interval $[T_{\rm fit},T]$. Again, the fit is performed by minimizing the mean squared error (RMSE) between the predicted and the simulated time-series,
where the average of $M = 1000$ microscopic simulations on a random $r$-regular graph with $r = 3$ and $N = 500$ nodes is used as the ground truth. We initialize the process by setting $1\%$ of the network nodes in the A state, and another $1\%$ of them in the I state.

\begin{figure}[t]
    \centering
    \includegraphics[width=\linewidth]{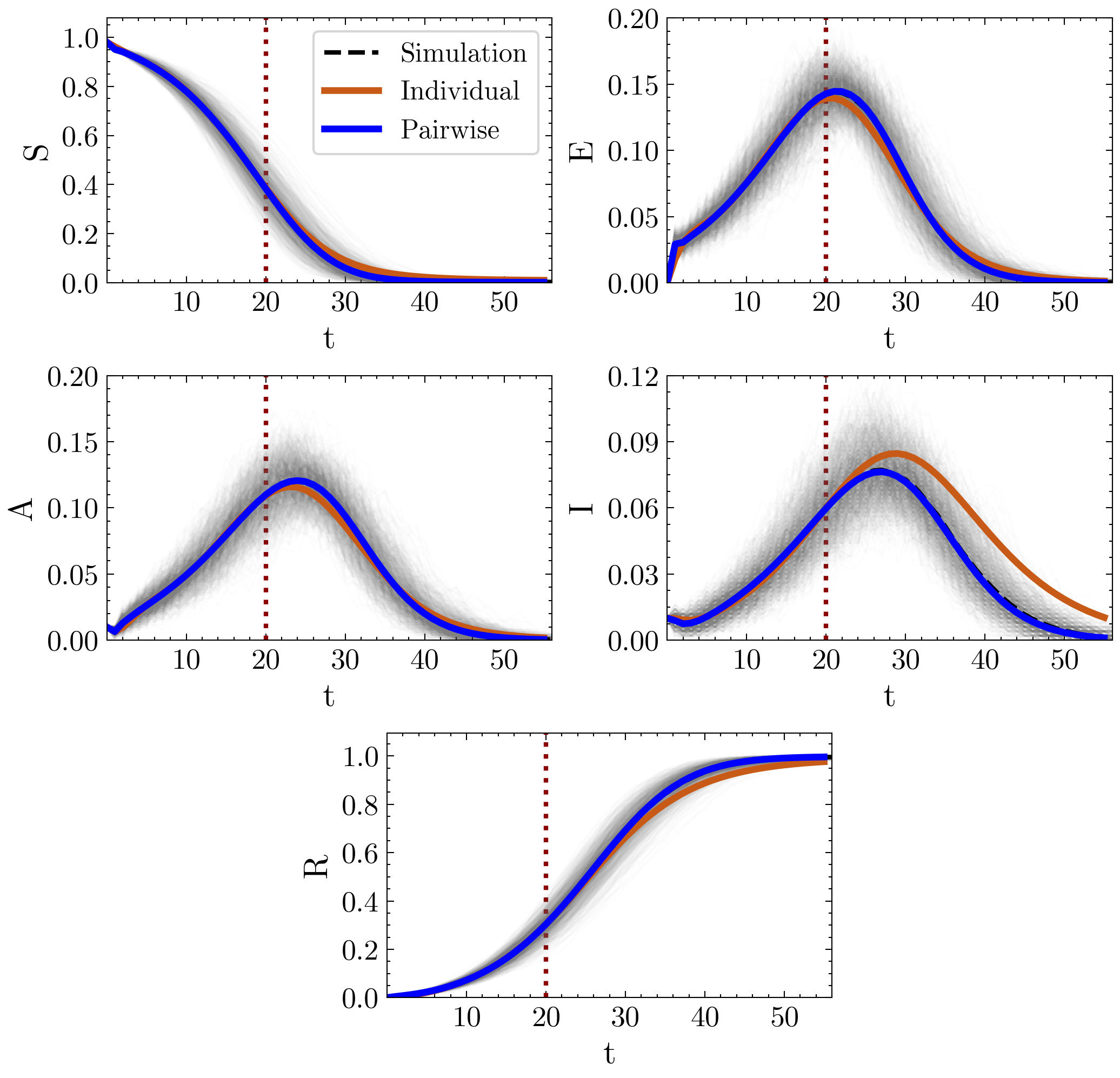}
    \caption{Comparison between stochastic simulations, individual-based model (orange solid line) and pair-based model (blue solid line) when all the state variables are fitted up to $T_{\rm fit}=20$ (red dotted line). The stochastic simulations are run on a random $r$-regular graph, with the same settings of Fig.~\ref{fig:reproducing_random_regular}. The stochastic simulations are represented as solid gray lines, while the average is plotted as a dashed black line.}
    \label{fig:fit_4B_t_20}
\end{figure}
As a first example, we fit the SEAIR models up to $T_{fin}=20$, and illustrate 
in Fig.~\ref{fig:fit_4B_t_20} 
the evolution of the state variables. Both the individual-based and the pairwise model reproduce quite well the temporal evolution of $\overline{S}_t$, $\overline{E}_t$, $\overline{A}_t$ and $\overline{R}_t$. However, as concerns the fraction symptomatic infectious $\overline{I}_t$, the individual-based model forecasting appears less reliable, while the pairwise model correctly predicts the time course of the variable. Moreover, if we consider the value of $\mathcal{D}(\mathbf{p})$, we can see that, in order to reproduce the system dynamics, the individual-based model has to rely on a set of epidemiological parameters that are substantially different from the one used to run the stochastic simulations. On the contrary, the pairwise model remains able to predict the correct set of parameters. Indeed, for the individual-based model we find $\mathcal{D}_{\rm ind}(\mathbf{p})=0.20$,  while we obtain $\mathcal{D}_{\rm pair}(\mathbf{p})=2.4\cdot10^{-3}$ for the pairwise model.

\begin{figure}[t]
    \centering
    \includegraphics[width=0.7\linewidth]{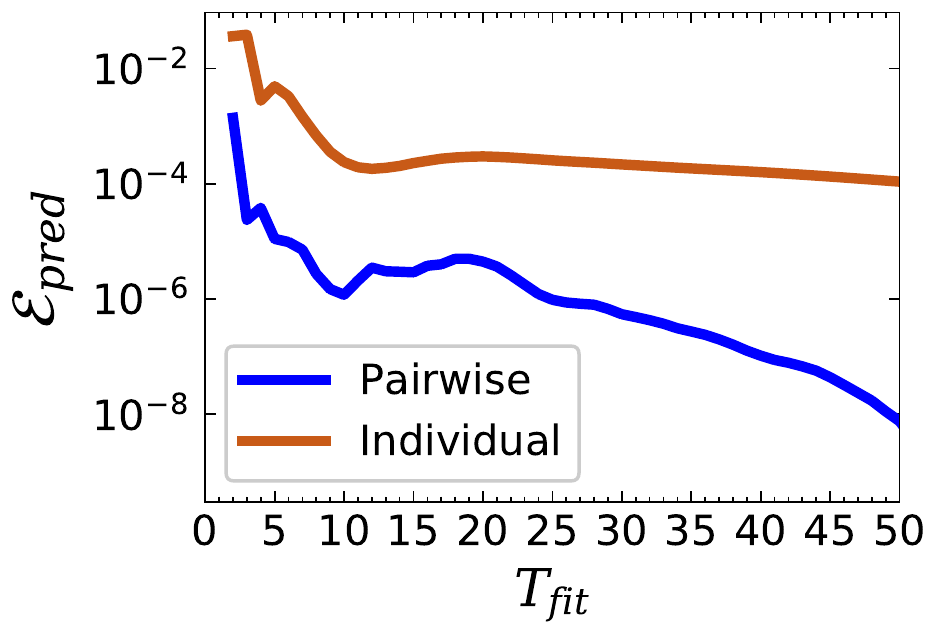}
    \caption{Value of the prediction error $\mathcal{E}_{\rm pred}$ as a function of the fitting range $T_{\rm fit}$, for both the individual-based (orange) and the pairwise (blue) models.}
    \label{fig:fit_MSE_t}
\end{figure}
In the previous example, we have compared the predictions of the deterministic models to the numerical simulations of the epidemic process by taking a single value of $T_{\rm fit}$ only. However, as the performance of the models may depend on the particular value of $T_{\rm fit}$ considered, a more systematic analysis is required. We do this by varying the value of the fitting time interval in the range $T_{\rm fit}\in [2,50]$, and monitoring the prediction error on the state variables in the time interval $[T_{\rm fit},T=55]$. As a measure of the model performance we consider the mean squared error between the predicted and the simulated time-series relative to all the state probabilities, namely
\begin{equation}
    \mathcal{E}_{\rm pred} = \sqrt{\frac{1}{|\Omega|(T-T_{\rm fit})}\sum_{X\in\Omega} \sum\limits_{t=T_{\rm fit}+1}^{T}( \langle X\rangle_t - \overline{X}_t )^2}\\ 
\end{equation}
Note that, since we consider a mean squared error, the value of $\mathcal{E}_{\rm pred}$ should not depend on the length of the time interval itself.  

Fig.~\ref{fig:fit_MSE_t} shows the results. Two aspects are worth remarking. First, the value of $\mathcal{E}_{\rm pred}$ relative to the pairwise model is generally lower compared to that of the individual-based model, meaning that the former provides more reliable predictions on the temporal evolution on the state variables. Second, the performance of the individual-based model is heavily influenced by the value of $T_{\rm fit}$, with the prediction error that decreases as the fitting time interval increases. Instead, the pairwise model provides a good prediction on the system dynamics even when it has at disposal only a limited amount of data, as one can see from the value of $\mathcal{E}_{\rm pred}$, which remains stable in the entire range of values of $T_{\rm fit}$.

In conclusion the pairwise model provides a more reliable forecasting of the temporal evolution of the state variables than the individual-based model.  
Furthermore, in those cases where the performances of the two models are comparable, the pairwise model yields a better prediction of the epidemiological parameters, consistently with the results obtained in the previous subsection.

\subsection{Estimating the unmeasurable variables}
Finally, we work under the worst condition (among those investigated), in which some of the dynamical variables are not measurable. At variance with the previous case study, where we have evaluated the epidemiological parameters assuming to be able to measure each compartment $X \in \Omega$, here we suppose to be able to measure only a subset of them,
fitting the deterministic models to the corresponding time series only. Understanding how the models perform under these circumstances is of crucial practical relevance in the context of outbreaks such as that of COVID-19, where the disease is also spread by asymptomatic carriers that are difficult to trace. 

In the analysis of the SEAIR models, we assume to be able to detect those individuals who are either symptomatic infectious (I) or recovered (R), considering unmeasurable the other states. We can consider two cases, which represent different degrees of data availability. In the first case, since empirical data are usually provided at an individual level, we assume the fractions of symptomatic and recovered individuals ($\overline{I}_t$, and $\overline{R}_t$) to be available data, while the fractions of susceptible, exposed and asymptomatic individuals ($\overline{S}_t$, $\overline{E}_t$, and $\overline{A}_t$) are unknown.
Notice that we are here assuming to be able to trace the asymptomatic individuals once they recover. This corresponds to the ideal case in which serological tests are systematically performed on the population, allowing to detect those individuals who have recovered without having been diagnosed with the disease.
Accordingly, to evaluate the epidemiological parameters, we fit the variables $\langle I\rangle_t$ and $\langle R\rangle_t$ to the corresponding time series. As discussed in Section \ref{section:epidemic_parameters}, there is a fundamental difference in the way in which the individual-based and the pairwise models are fitted. Indeed, while in the individual-based model the probabilities $\langle X \rangle_t$ are the state variables, in the pairwise model these quantities are derived from the pair probabilities through relation \eqref{eq:marginal}. This means that, at variance with the individual-based model, for the pairwise model a function of the state variables is used in the data fitting.

In the second case, in addition to $\overline{I}_t$ and $\overline{R}_t$, we assume to have a larger set of measured time series, including, in particular, some of the pair variables. Indeed, when we have information on the contact network structure, if we are able to detect those individuals that are either in state I or in state R, we can also measure the pairs formed by individuals in one of these two states. To account for this scenario, we have considered the fraction of pairs composed either by symptomatic or recovered individuals as measurable variables, i.e. 
$\overline{II}_t$, $\overline{IR}_t$, and $\overline{RR}_t$, while the remaining fractions are considered unknown (as implicitly assumed before). Under these hypotheses, for the pairwise model these data do not have to be fitted as a function of the state variables, since they can be directly associated to the corresponding (measurable) pair variables $\langle II\rangle_t$, $\langle IR\rangle_t$, and $\langle RR\rangle_t$. Summing up, in this second case, we fit the individual-based model to $\overline{I}_t$ and $\overline{R}_t$, and the pairwise model to $\overline{I}_t$, $\overline{R}_t$, $\overline{II}_t$, $\overline{IR}_t$ and $\overline{RR}_t$.

\begin{figure}[t]
    \centering
    \includegraphics[width=\linewidth]{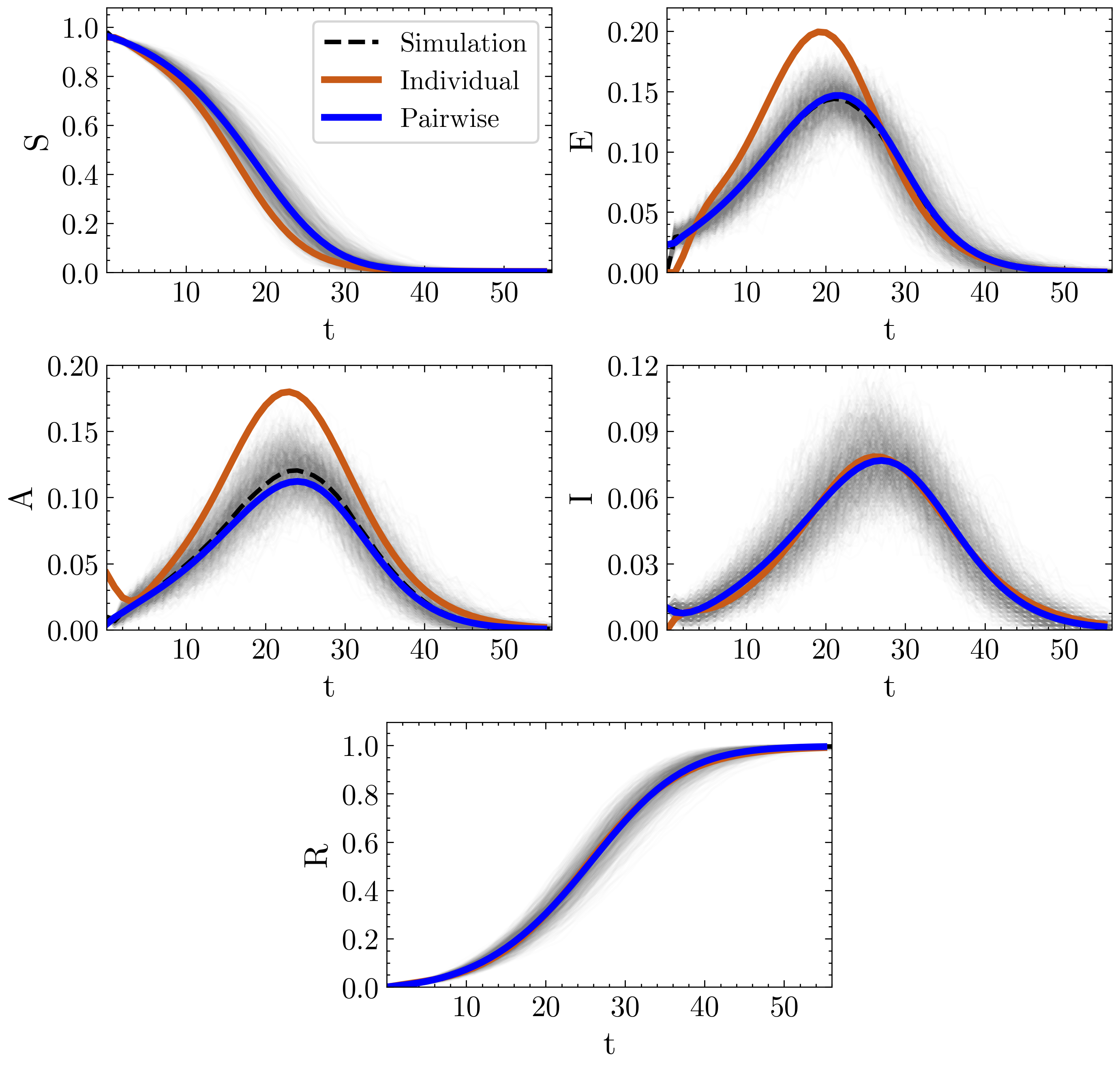}
    \caption{Comparison between stochastic simulations, individual-based model (orange solid line) and pair-based model (blue solid line) when the measurable state variables are fitted over the entire time course of the epidemic process (i.e. $T_{\rm fit}=T$). The stochastic simulations are run on a random $r$-regular graph, with the same settings of Fig.~\ref{fig:reproducing_random_regular}, and their results are represented as solid gray lines, while the average is plotted as a dashed black line.}
    \label{fig:fit_measurable_all_TS_t56}
\end{figure}

\begin{figure}[t]
    \centering
    \includegraphics[width=\linewidth]{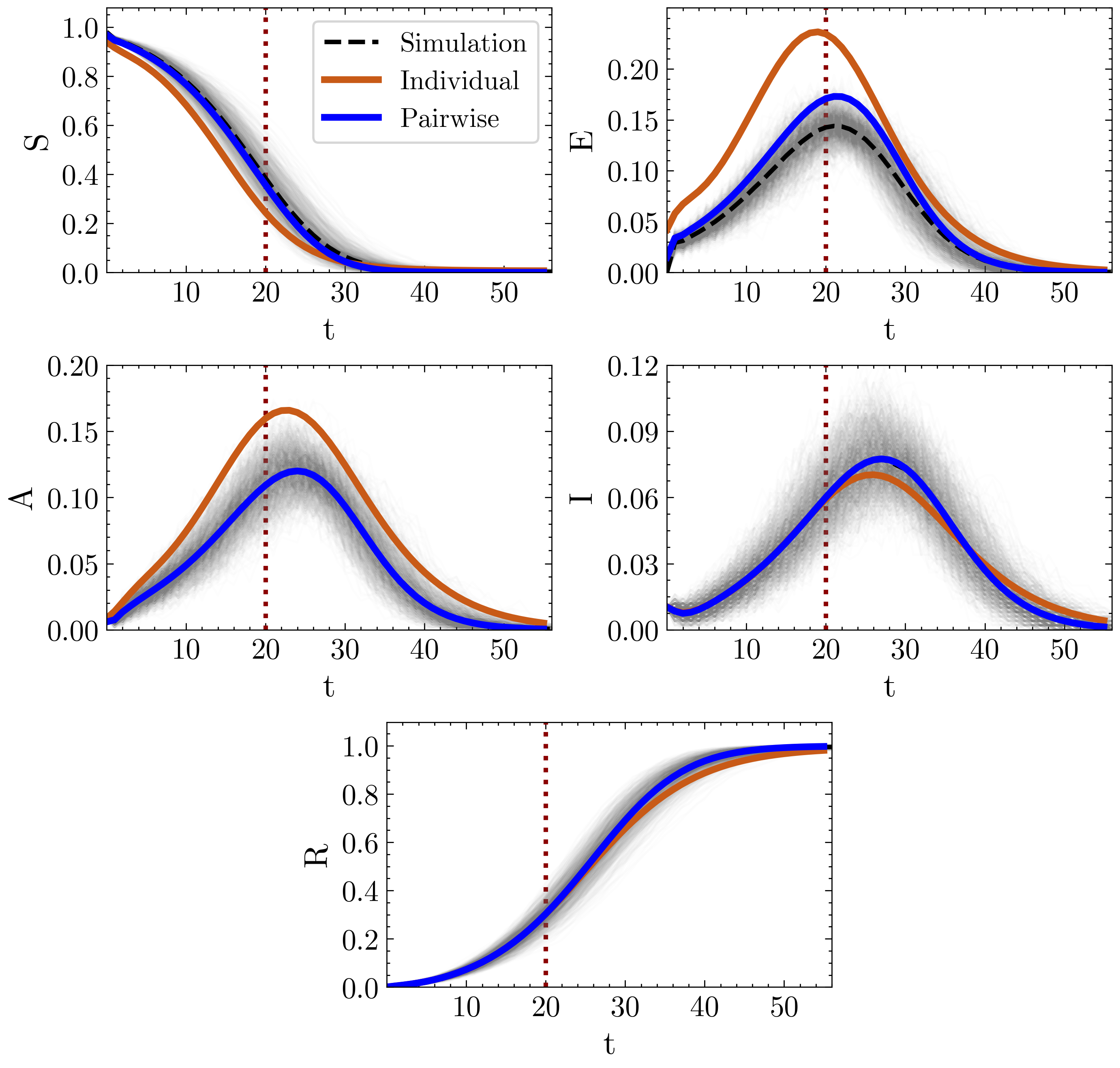}
    \caption{Comparison between stochastic simulations, individual-based model (orange solid line) and pair-based model (blue solid line) when the measurable state variables are fitted up to $T_{\rm fit}=20$ (red dotted line). The stochastic simulations are run on a random $r$-regular graph, with the same settings of Fig.~\ref{fig:reproducing_random_regular}, and their results are represented as solid gray lines, while the average is plotted as a dashed black line.}
    \label{fig:fit_measurable_all_TS_t20}
\end{figure}

We begin our analysis from this latter case, which relies on the assumption that the contact network structure is known. Then we discuss the implications of fitting the pairwise model when such data are not available.

The dynamics of the SEAIR models is ultimately determined by two sets of quantities, namely the epidemiological parameters $\mathbf{p}$ and the initial condition of the state variables. Since, in the previous sections, we have assumed all the dynamical variables to be known at any time, we have considered the initial conditions as known quantities, fitting the SEAIR models to empirical data only to evaluate $\mathbf{p}$.
Now, as some of the variables are unknown, also their initial conditions need to be estimated from the fit. It is worth noting that the pairwise model has a larger number of unknown initial conditions compared to the individual-based model. This means that, at variance with the cases in Sections \ref{section:epidemic_parameters} and \ref{section:predictive_capability}, where the number of unknown parameters was equal for both models, in the present scenario one needs to determine a larger set of parameters for the pairwise model.

We start by considering Fig.~\ref{fig:fit_measurable_all_TS_t56} and Fig.~\ref{fig:fit_measurable_all_TS_t20}, in which we compare the prediction of the individual-based and of the pairwise model to the stochastic simulations, for two different values of $T_{\rm fit}$. As in the previous case, the numerical data are generated performing $M=1000$ microscopic simulations on a random $r$-regular graph with $N=500$ and $r=3$, and taking the average dynamics as the ground truth. First, we discuss the case $T_{\rm fit}=T$, that represents an important case study occurring in the analysis of past epidemic outbreaks. In this scenario, we assume we know the dynamics of the measurable variables $\overline{I}_t$ and $\overline{R}_t$ over the entire time course of the epidemic, while the other variables remain unknown. In other words, we here analyze the capability of the models to determine the temporal evolution of the unmeasurable variables once the epidemic outbreak is over. As shown in Fig.~\ref{fig:fit_measurable_all_TS_t56}, the pairwise model is in a much better agreement with the stochastic simulations compared to the individual-based one.

Conversely, to illustrate an example in which data is only available for a limited time interval, we consider the case $T_{\rm fit}=20$. The results for this case study are displayed in Fig.~\ref{fig:fit_measurable_all_TS_t20}. First, we note that, compared to the individual-based model, the pairwise model better forecasts the temporal evolution of the measurable variables $\overline{I}_t$ and $\overline{R}_t$. Second, the pairwise model provides a good prediction of the time course of the unknown variables. In particular, we remark that the pairwise model prediction of both $\overline{A}_t$ and $\overline{S}_t$ is in good agreement with the numerical simulations. As regards the probability of being exposed $\langle E\rangle_t$, though it overestimates the average behavior of the stochastic simulations $\overline{E}_t$, we note that time series predicted by the pairwise model represents a significant improvement compared to the prediction of the individual-based model. Overall, the pairwise model provides a reliable prediction of the dynamics of both measured and unmeasured variables, outperforming the individual-based model.

\begin{figure}[t]
    \centering
    \includegraphics[width=0.7\linewidth]{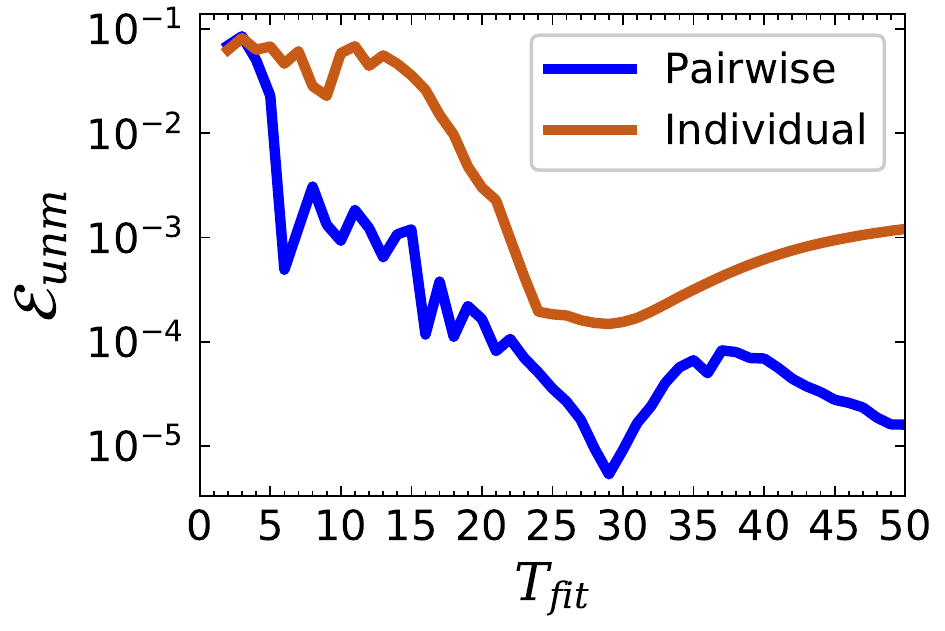}
    \caption{Value of the prediction error $\mathcal{E}_{unm}$ as a function of the fitting range $T_{\rm fit}$, for both the individual-based (orange) and the pairwise (blue) models.}
    \label{fig:fit_measurable_RMPE}
\end{figure}

Then, we study the predictive capability of the deterministic models as function of $T_{\rm fit}$. Fig.~\ref{fig:fit_measurable_RMPE} displays, for both the individual-based and the pairwise models, the prediction error $\mathcal{E}_{\rm unm}$ on the unmeasured individual-level time series, namely 
\begin{equation}
    \mathcal{E}_{\rm unm} =\sqrt{ \frac{1}{|\Theta|(T+1)}\sum_{X\in\Theta} \sum\limits_{t=0}^{T}( \langle X\rangle_t - \overline{X}_t )^2}\\ 
\end{equation} 
where $\Theta = \Omega \setminus \Omega' = \{S , E, A\}$ represents the set of unmeasured compartments. Note that, since we are interested in the prediction over the entire course of the epidemic, we considered the range $[0,T]$.

Consistently with the previous results, we note that, for different values of $T_{\rm fit}$, the pairwise model generally provides a lower value of $\mathcal{E}_{\rm unm}$ compared to individual-based model, meaning that the former furnishes a more reliable prediction of the temporal evolution of the unknown variables. 

\begin{figure}[t]
    \centering
    \includegraphics[width=0.7\linewidth]{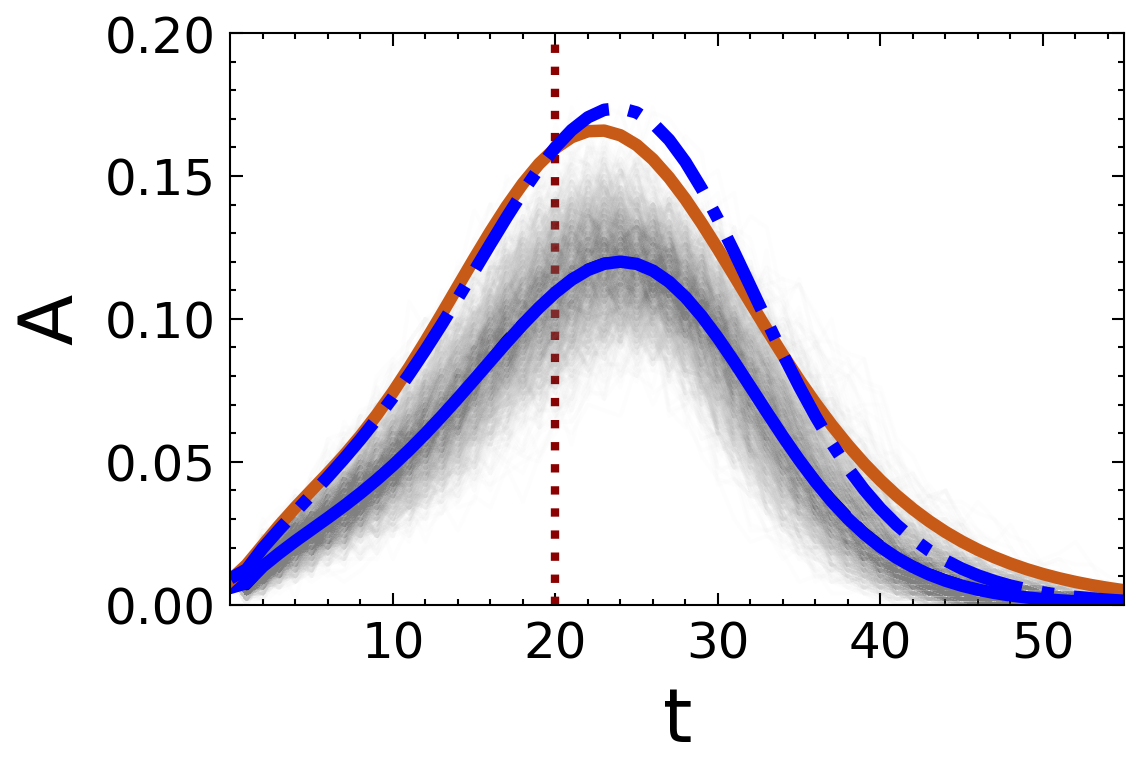}
    \caption{Comparison between stochastic simulations (dashed black), individual-based model (solid orange) and pairwise model for the prediction of  the asymptomatic state of Fig.~\ref{fig:fit_measurable_all_TS_t20}. The pairwise model has been fitted both including (solid blue) and excluding (dashed-dotted blue) the time series $\overline{II}_t$, $\overline{IR}_t$ and $\overline{RR}_t$ from the set of measurable variables. Note that the solid blue line is superimposed to the dashed black one.}
    \label{fig:fit_measurable_comparison_pairwise_on_A}
\end{figure}

The results discussed so far indicate that the pairwise model outperforms the individual-based model in predicting the dynamics of the fractions of individuals in the unmeasured compartments. However, in order to obtain a more reliable prediction on quantities that are difficult or even impossible to measure, e.g.\ the asymptomatic infectious population, it is necessary to gather more and different data about the quantities that are accessible. In particular, here we assumed we are able to measure not only the fraction of individuals that are symptomatic infectious or recovered, but also the fraction of couples formed by individuals in one of those states.

To conclude, we discuss how important these data are for the prediction, by showing what happens when they are not available. Fig.~\ref{fig:fit_measurable_comparison_pairwise_on_A} displays the prediction of the fraction of asymptomatic infectious $\overline{A}_t$ (which we assumed to be unmeasurable) by the individual-based and pairwise model. While the individual-based model is fitted only to $\overline{I}_t$ and $\overline{R}_t$, the pairwise model is fitted either using or not the fractions of pairs $\overline{IR}_t$, $\overline{IR}_t$ and $\overline{RR}_t$. As it can be noticed, when only the individual probabilities are fitted, the pairwise model is not able to provide a good prediction for $\overline{A}_t$, performing as badly as the individual-based model. 

Overall, the results of this section suggest that using a model that accounts for the dynamical correlations existing within the contact network can ensure a more accurate evaluation of the unknown quantities of the system. However, this approach demands to gather more refined data on how the disease spreads through the network itself, using, for instance, more detailed contact tracing techniques \cite{eames2003contact}.

\section{Conclusions \label{section:conclusion}}
In this work, we have developed two discrete-time population-level SEAIR models, providing a coarse grained description of the spreading of an infectious disease through a network of contacts. First, under the hypothesis of statistical independence at the level of nodes, we have derived the master equations for an individual-based model. Second, we have considered a more complex pairwise approximation, describing the system at the level of node pairs. This allowed us to account for the dynamical correlations existing within the contact network. 

Assuming we know the epidemiological parameters characterizing the disease spreading, we have compared the deterministic models to numerical simulations of the epidemic dynamics. We have analyzed to which degree the predictions of the SEAIR models agreed with the results of stochastic simulations carried out over two different network topologies, namely the random $r$-regular graph and the Erd\"os-Renyi graph. Consistently with previous research \cite{frasca2016discrete}, we found that the pairwise model is able to reproduce, for both topologies, the temporal evolution of the state variables, whereas the individual-based model overestimates the number of new infections occurring at each time step. This result was consistently observed when important parameters of the model, such as the symptomatic and asymptomatic transmission probabilities, were varied.

We have then considered the more realistic case in which the epidemiological parameters are unknown and need to be estimated by fitting the SEAIR models to empirical data. We have examined three case studies of increasing complexity. First, we have assumed to know the time-course of all the state variables for the entire duration of the epidemic outbreak, analyzing the capability of the deterministic models to estimate the epidemiological parameters. Compared to the individual-based model, the pairwise model performed better, as the error on the parameters resulted generally lower and independent on their specific value chosen for the numerical simulations.
This aspect is crucial in those practical situations where one needs to know the exact value of the parameters, for instance to evaluate the possible effects of issuing/lifting a containment policy, which can be assumed to have an impact only on specific parameters (for instance, the obligation to wear a face mask likely affects only the transmission probabilities $\beta_A$ and $\beta_I$).

Second, we have assumed to know the dynamics of the state variables for a limited time interval only, analyzing the capability of both the individual and the pairwise model to forecast the evolution of the pandemic. Our results show that the individual-based model is not able to correctly forecast the dynamics of all the state variables, while the prediction of the pairwise model is in good agreement with the numerical simulations. Furthermore, the pairwise model still provides a better estimate of the epidemiological parameters even under these conditions. This higher reliability of the pairwise approach can play a fundamental role in the policy-making process, where accurately forecasting the evolution of the epidemic is crucial.

Third, we considered a more realistic scenario, assuming we are able to measure only a subset of the model compartments. In particular, we have assumed to know the fractions of symptomatic infected and recovered nodes, i.e. $\overline{I}_t$ and $\overline{R}_t$. Additionally, we have considered to be able to measure the pairs in which the nodes are either in the infectious symptomatic or in the recovered states, i.e. $\overline{II}_t$, $\overline{IR}_t$ and $\overline{RR}_t$. With this information available, the fractions of individuals in the unmeasured compartments, i.e. $\overline{S}_t$, $\overline{E}_t$ and $\overline{A}_t$, as predicted by the pairwise model were found in good agreement with the numerical simulations. On the contrary, the individual-based model was not able to estimate their temporal evolution. Since the asymptomatic individuals can constitute a public-health risk, as they are able to spread the virus without knowing it, and since they can be difficult to trace, assessing their number within the population becomes crucial, and the pairwise approach can provide an important instrument to cope with their presence.

Overall, the results presented in this paper show that the pairwise modeling paradigm is a reliable tool for estimating the epidemiological parameters and forecasting the evolution of the epidemics. 
In particular, we showed that a pairwise approach, altogether with additional information regarding the contacts of the infectious individuals, provides a much better prediction of the unmeasurable variables of the system compared to an individual-based one.
Moreover, such a modeling approach allows one to evaluate the size of the asymptomatic population, which, in the context of the COVID-19 pandemic, represents a critical issue. To fulfill this objective, the pairwise model requires to refine the data collection procedures so as to include information on the network structure. For this reason, extending the use of contact tracing techniques can play a crucial role in informing mathematical modeling, thus leading to more reliable predictions on the course of epidemics.


Finally, we note that the generality of the approach here discussed paves the way to applications to other classes of epidemic models (possibly including other compartments or an explicit dependency of the contact networks on time), which are possible directions for future work.

\section*{Acknowledgments}
V.L and G.R. acknowledge support from University of Catania project ``Piano della Ricerca 2020/2022, Linea d’intervento 2, MOSCOVID''. G.R.has been partially supported by the Italian Ministry of Instruction, University and Research (MIUR) through PRIN Project 2017, No. 2017KKJP4X.

\section*{Appendix A: Derivation of Eq.~(\ref{eq:PiSEchiusaInd})}
\label{App_A}
\setcounter{equation}{0}
\renewcommand{\theequation}{A\arabic{equation}}

Here we show how to obtain the expression for the nonlinear transition probability $\Pi_{SE\rightarrow EA}$ of the pairwise SEAIR model, i.e. Eq.~(\ref{eq:PiSEchiusaInd}). To start with, 
we introduce the homogeneous mixing hypothesis to write Eq.~(\ref{eq:transition_individual}) at the level of population, thus dropping the indices in the probability terms. 
This yields
\begin{equation}
    \label{eq:transition_individual_population}
    \begin{array}{l}
    \Pi_{S \rightarrow E} \approx \langle SIU\ldots U \rangle_t [1-(1-\beta_I)] \\
    + \langle SUI\ldots U \rangle_t [1-(1-\beta_I)] \\
     +\ldots\\[3pt]
    + \langle SUU\ldots I \rangle_t [1-(1-\beta_I)] \\
    + \langle SAU\ldots U \rangle_t [1-(1-\beta_A)] \\
    +\ldots\\[3pt]
    + \langle SUU\ldots A \rangle_t [1-(1-\beta_A)] \\
    + \langle SII\ldots U \rangle_t [1-(1-\beta_I)^2]  \\
    +\ldots\\[3pt]
    + \langle SIU\ldots I \rangle_t [1-(1-\beta_I)^2] \\
    + \langle SIA\ldots U \rangle_t [1-(1-\beta_I)(1-\beta_A)]\\
    +\ldots\\
    + \langle SAA\ldots A \rangle_t [1-(1-\beta_A)^k]. 
    \end{array}
\end{equation}

As explained in Section \ref{sub:ib}, given the expression in Eq.~(\ref{eq:transition_individual_population}), the system~(\ref{eq:individual}) is not closed.
To close it at the level of individuals, we consider the approximation in Eq.~(\ref{eq:individual_closure}), which can be also rewritten at a population-level by dropping the node indices. By substituting this expression in Eq.~(\ref{eq:transition_individual_population}), we get

\begin{equation}
    \begin{array}{l}
    \Pi_{S \rightarrow E} \approx \langle S\rangle_t \langle I \rangle_t \langle U \rangle_t \ldots \langle U \rangle_t [1-(1-\beta_I)] \\[3pt]
    + \langle S\rangle_t \langle U \rangle_t \langle I \rangle_t \ldots \langle U \rangle_t [1-(1-\beta_I)] \\[3pt]
     +\ldots \\[3pt]
    + \langle S\rangle_t \langle U \rangle_t \langle U \rangle_t \ldots \langle I \rangle_t [1-(1-\beta_I)] \\[3pt]
    + \langle S\rangle_t \langle A \rangle_t \langle U \rangle_t \ldots \langle U \rangle_t [1-(1-\beta_A)] \\[3pt]
    +\ldots \\[3pt]
    + \langle S\rangle_t \langle U \rangle_t \langle U \rangle_t \ldots \langle A \rangle_t [1-(1-\beta_A)] \\[3pt]
    + \langle S\rangle_t \langle I \rangle_t \langle I \rangle_t \ldots \langle U \rangle_t [1-(1-\beta_I)^2] \\[3pt]
    +\ldots \\[3pt]
    + \langle S\rangle_t \langle I \rangle_t \langle U \rangle_t \ldots \langle I \rangle_t [1-(1-\beta_I)^2] \\[3pt]
    + \langle S\rangle_t \langle I \rangle_t \langle A \rangle_t \ldots \langle U \rangle_t [1-(1-\beta_I)(1-\beta_A)] \\[3pt]
    +\ldots \\[3pt]
    + \langle S\rangle_t \langle A \rangle_t \langle A \rangle_t \ldots \langle A \rangle_t [1-(1-\beta_A)^k], 
    \end{array}
\end{equation}

\noindent which can rewritten as
\begin{equation}
    \begin{array}{l}
    \Pi_{S \rightarrow E} \approx {k \choose 1}{1 \choose 0}\langle S\rangle_t \langle I \rangle_t \langle U \rangle_t^{k-1} [1-(1-\beta_I)] \\[3pt]
    + {k \choose 1}{1 \choose 1}\langle S\rangle_t \langle A \rangle_t \langle U \rangle_t^{k-1} [1-(1-\beta_A)] \\[3pt]
    + {k \choose 2}{2 \choose 0}\langle S\rangle_t \langle I \rangle_t^2 \langle U \rangle_t^{k-2} [1-(1-\beta_I)^2] \\[3pt]
    + {k \choose 2}{2 \choose 1}\langle S\rangle_t \langle I \rangle_t \langle A \rangle_t \langle U \rangle_t^{k-2} [1-(1-\beta_I)(1-\beta_A)] \\[3pt]
    +\ldots \\[3pt]
    + {k \choose k}{k \choose k}\langle S\rangle_t \langle A \rangle_t^{k}[1-(1-\beta_A)^k]. 
    \end{array}
\end{equation}
Note that each term is characterized by the product of two binomial factors, i.e. ${k \choose p}{p \choose n}$. The first one indicates the number of ways $p$ infected nodes can be chosen among the $k$ neighbors of a node, while the second corresponds to the number of ways $n$ asymptomatic infectious nodes can be chosen among the $p$ infected neighbors. More compactly, we can write
\begin{equation}
\label{eq:compact_sum_individual_closed}
    \begin{array}{l}
    \Pi_{S \rightarrow E} \approx \langle S\rangle_t \sum \limits_{p=1}^{k}\sum\limits_{n=0}^{p}{k \choose p}{p \choose n} \langle A \rangle_t^{n}\langle I \rangle_t^{p-n} \langle U \rangle_t^{k-p}\\[10pt]
    \times\left[1-(1-\beta_A)^{n}(1-\beta_I)^{p-n} \right]. \\[5pt]
\end{array}
\end{equation}

Given that $\langle U \rangle_t = 1 - \langle A \rangle_t - \langle I \rangle_t$, it is worth noting that the term ${k \choose p}{p \choose n} \langle A \rangle_t^{n}\langle I \rangle_t^{p-n} \langle U \rangle_t^{k-p}$ corresponds to the probability mass function of the multinomial distribution (see \cite{Ross_Probability_2010}, Chapter 6). With this in mind, we can manipulate Eq.~(\ref{eq:compact_sum_individual_closed}) so to have
\begin{equation}
\label{eq:pre_multionomial}
    \begin{array}{l}
    \Pi_{S \rightarrow E} \approx \langle S\rangle_t[1 - \sum \limits_{p=0}^{k}\sum\limits_{n=0}^{p}{k \choose p}{p \choose n} \langle A \rangle_t^{n}\langle I \rangle_t^{p-n}\langle U\rangle_t^{k-p}\\[10pt]
    \times \left(1-\beta_A\right)^{n}\left(1-\beta_I\right)^{p-n}] \\[10pt]
    = \langle S\rangle_t\{1 - \sum \limits_{p=0}^{k}\sum\limits_{n=0}^{p}{k \choose p}{p \choose n} \langle U\rangle_t^{k-p}\\[10pt]
    \times[\langle A \rangle_t(1-\beta_A)]^{n}[\langle I \rangle_t(1-\beta_I)]^{p-n}\}\\[10pt]
    = \langle S\rangle_t\left[1 - \sum \limits_{p=0}^{k}\sum\limits_{n=0}^{p}\frac{k!}{n!(p-n)!(k-p)!} x_1^{k-p}x_2^n x_3^{p-n}\right],
    \end{array}
\end{equation}
where we have defined $x_1 = \langle U\rangle_t$, $x_2=\langle A\rangle_t(1-\beta_A)$ and $x_3=\langle I\rangle_t(1-\beta_I)$.
Note also that we added to the summation the term relative to $p=0$, since this is equal to zero, and that we have considered that summing the probability mass function over all the possible values of $p$ and $n$ gives one. 
By recalling the multinomial theorem
\begin{equation}
    (x_1 + x_2 + x_3)^{k} = \sum\limits_{n_1+n_2+n_3=k}\frac{k!}{n_1!n_2!n_3!}\prod\limits_{i=1}^{3} x_i^{n_i},
    \label{eq:multinomial_theorem}
\end{equation}
we can simplify Eq.~(\ref{eq:pre_multionomial}), obtaining
\begin{equation}
    \begin{array}{l}
    \Pi_{S\rightarrow E} \approx \langle S \rangle_t [1 - (x_1+x_2+x_3)^{k}].
    \end{array} 
\end{equation}
Finally, by substituting $x_1+x_2+x_3 = 1 - \beta_A\langle A\rangle_t - \beta_I\langle I\rangle_t$ (note again that $\langle U\rangle_t = 1 - \langle A\rangle_t - \langle I\rangle_t$), we obtain for $\Pi_{S\rightarrow E}$ the expression in Eq.~(\ref{eq:PiSEchiusaInd}).

\section*{Appendix B: Derivation of Eq.~(\ref{eq:pairwise_population_SE_EA})}
\label{App_B}
\setcounter{equation}{0}
\renewcommand{\theequation}{B\arabic{equation}}
Here we derive the expression for the nonlinear transition probability $\Pi_{SE\rightarrow EA}$ of the pairwise SEAIR model, i.e. Eq.~(\ref{eq:pairwise_population_SE_EA}). We begin by considering Eq.~(\ref{eq:transition_pair}), introducing the assumption that no triangular loops exist within the network and that the population is homogeneously mixed. In this case, all nodes have the same number of neighbors, $k$, so that the number of neighbors of a pair of nodes is $L=2k-2$. A graphical representation of the subgraph induced by the pair in state $(S,E)$ under these hypotheses is provided in Fig.~\ref{fig:subgraph_pairwise}. Eq.~\eqref{eq:transition_pair} can written as

\begin{equation}
    \begin{array}{ll}
    &\Pi_{SE \rightarrow EA} \approx \langle SEIU\ldots U \rangle_t [1-(1-\beta_I)]\alpha_{EA} \\
    &+ \langle SEUI\ldots U \rangle_t [1-(1-\beta_I)]\alpha_{EA}\\
    & +\ldots\\
    &+ \langle SE\underbrace{U\ldots UI}_{k-1}\underbrace{U \ldots U}_{k-1}\rangle_t [1-(1-\beta_I)]\alpha_{EA}\\
    &+ \langle SEA\ldots U \rangle_t [1-(1-\beta_A)]\alpha_{EA}\\[3pt]
    &+\ldots\\
    &+ \langle SEU\ldots UAU \ldots U\rangle_t [1-(1-\beta_A)]\alpha_{EA}\\
    &+ \langle SEII\ldots U \rangle_t [1-(1-\beta_I)^2]\alpha_{EA}\\
    &+\ldots\\
    &+ \langle SEIU\ldots I \rangle_t
    [1-(1-\beta_I)]\alpha_{EA}\\
    &+ \langle SEIA\ldots U \rangle_t [1-(1-\beta_I)(1-\beta_A)]\alpha_{EA}\\
    &+\ldots\\
    &+ \langle SEAA\ldots A \rangle_t [1-(1-\beta_A)^{k-1}]\alpha_{EA}.
    \end{array}
    \label{eq:transition_pair_population}
\end{equation}
Each term $\langle SE\ldots Z\rangle_t$ counts $L+2 = 2k$ elements. The first two, i.e. S and E, represent the state of the pair of nodes inducing the subgraph, the following $k-1$ indicate the state of the neighborhood of the first node (the one in state S), while the remaining $k-1$ denote the state of the neighborhood of the second node (in state E). 
Eq.~\eqref{eq:transition_pair_population} can be simplified taking into account the property of symmetry of the joint probabilities. Indeed, since under the homogeneous mixing hypothesis the nodes of the neighborhood are statistically equivalent to one another, the probability terms only depend on the number of infected nodes in each neighborhood, while they do not depend on which particular node is infected. For example, when only one symptomatic infectious node is present in the subgraph, assuming it to be connected to the node in state S, we have $k-1$ equivalent terms, namely
\begin{equation}
\begin{array}{l}
\langle SEIUU\ldots U \rangle_t = \langle SEUIU\ldots U \rangle_t = \ldots \\
\ldots = \langle SE\underbrace{UU\ldots UI}_{k-1}U \ldots U\rangle_t.
\end{array}
\end{equation}
Therefore, we can rewrite Eq.~\eqref{eq:transition_pair_population} as
\begin{equation}
    \begin{array}{ll}
    &\Pi_{SE \rightarrow EA} \approx \\[3pt]
    &\langle SEIUU\ldots U \rangle_t{k-1 \choose 1}{1 \choose 0}{k-1 \choose 0}{0 \choose 0}[1-(1-\beta_I)]\alpha_{EA} \\[3pt]
    &+ \langle SEAU\ldots U \rangle_t {k-1 \choose 1}{1 \choose 0}{k-1 \choose 0}{0 \choose 0} [1-(1-\beta_A)]\alpha_{EA}\\[3pt]
    &+ \langle SEII\ldots U \rangle_t {k-1 \choose 2}{2 \choose 0}{k-1 \choose 0}{0 \choose 0} [1-(1-\beta_I)^2]\alpha_{EA}\\[3pt]
    &+ \langle SEIU\ldots I \rangle_t {k-1 \choose 1}{1 \choose 0}{k-1 \choose 1}{1 \choose 0}
    [1-(1-\beta_I)]\alpha_{EA}\\[3pt]
    &+ \langle SEIA\ldots U \rangle_t {k-1 \choose 2}{2 \choose 1}{k-1 \choose 0}{0 \choose 0} \\[3pt]
    &\times[1-(1-\beta_I)(1-\beta_A)]\alpha_{EA}\\[3pt]
    &+\ldots\\[3pt]
    &+ \langle SEAA\ldots A \rangle_t {k-1 \choose k-1}{k-1 \choose k-1}{k-1 \choose k-1}{k-1 \choose k-1} \\[3pt] &\times[1-(1-\beta_A)^{k}]\alpha_{EA}.
    \end{array}
    \label{eq:transition_pair_population_symmetry}
\end{equation}
We note that each term is characterized by the product of four binomial factors, i.e. ${k-1 \choose p}{p \choose n}{k-1 \choose q}{q \choose m}$, representing the number of possible combinations of the neighboring nodes. The first two binomial coefficients are relative to the neighborhood of the node in state S, while the others concern the neighbors of the node in state E. For each of them, the first binomial factor indicates the number of ways $p$ (or $q$) infected nodes can be chosen among the $k-1$ neighbors of a node, while the second corresponds to the number of ways $n$ (or $m$) asymptomatic infectious nodes can be chosen among the $p$ (or $q$) infected neighbors.

As explained in Section \ref{sec:pair_model}, given the expression in Eq.~(\ref{eq:transition_pair_population}), the system~(\ref{eq:pair}) is not closed. To close it at the level of pairs, we consider the approximation in Eq.~(\ref{eq:node_level_closure}), which can be also rewritten at a population-level by dropping the node indices. With reference to the configuration of Fig.~\ref{fig:subgraph_pairwise}, we  can approximate the probability that the node S is connected to $n_I$ neighbors in state I, to $n_A$ neighbors in state A, and to $k-1-n_I-n_A$ neighbors in state $U$, while the node in state E is connected to $m_I$ neighbors in state I, to $m_A$ neighbors in state A, and to $k-1-m_I-m_A$ neighbors in state $U$, as follows
\begin{equation}
\begin{array}{l}
F(\langle SE \rangle_t,\ldots,\langle EA\rangle_t) = \langle SE\rangle_t\\[5pt]
\times \frac{\langle SA\rangle_t^{n_A}\langle SI\rangle_t^{n_I}\langle SU\rangle_t^{k-1-n_A-n_I}}{\langle S \rangle_t^{k-1}}\\[5pt]
\times\frac{\langle EA\rangle_t^{m_A}\langle EI\rangle_t^{m_I}\langle EU\rangle_t^{k-1-m_A-m_I}}{\langle E \rangle_t^{k-1}}.
\end{array}
\label{eq:pop_level_closure}
\end{equation}
By substituting the closure~(\ref{eq:pop_level_closure}) in Eq.~(\ref{eq:transition_pair_population_symmetry}), we obtain the expression for the transition probability in Eq.~(\ref{eq:double_sum}). As the transition of an individual from state E to state A is independent on the state of the neighbors, we expect the transition probability $\Pi_{SE\rightarrow EA}$ to be independent of the state probabilities $\langle EX\rangle$. Consistently, we note that the second double summation term, which accounts for all the possible configurations of the neighborhood of the node in state E, sums to one. To show this, we first define the quantities
\begin{equation}
    \varepsilon_A = \frac{\langle EA\rangle_t}{\langle E\rangle_t}, \qquad \varepsilon_I = \frac{\langle EI\rangle_t}{\langle E\rangle_t}, \qquad \varepsilon_U = \frac{\langle EU\rangle_t}{\langle E\rangle_t},
\end{equation}
which represent the probability that a node in state E is connected either to a node in state A, I or U, respectively, divided by the probability of being in state E.
With this new notation, the relation on the marginal probabilities $\langle E \rangle_t = \langle EA \rangle_t + \langle EI \rangle_t + \langle EU \rangle_t$ now reads
\begin{equation}
   \varepsilon_A + \varepsilon_I + \varepsilon_U = 1.
\end{equation}
Note that, despite it is not explicitly indicated, the terms $\varepsilon_A$, $\varepsilon_I$ and $\varepsilon_U$ clearly depend on time. 

Each term of the second double summation in Eq.~(\ref{eq:double_sum}) can now be rewritten as
\begin{equation}
\begin{array}{l}
\frac{\langle EA \rangle_t^{m}\langle EI \rangle_t^{q-m}\langle EU \rangle_t^{k-1-q}}{\langle E\rangle_t^{k-1}}{k-1 \choose q}{q \choose m} = \\[10pt]
\varepsilon_A^m \varepsilon_I^{q-m}\varepsilon_U^{k-1-q}\frac{(k-1)!}{m!(q-m)!(k-1-q)!},
\label{eq:multinomial_distr}
\end{array}
\end{equation}
which corresponds to the probability mass function associated to the multinomial distribution. Since we sum over all the possible values of $q$ and $m$, the second summation in Eq.~(\ref{eq:double_sum}), as we expect, is equal to one. We can thus rewrite the equation as
\begin{equation}
    \begin{array}{l}
    \Pi_{SE\rightarrow EA} \approx \langle SE \rangle_t \sum\limits_{p=1}^{k-1}\sum\limits_{n=0}^{p} \frac{\langle SA \rangle_t^{n}\langle SI \rangle_t^{p-n}\langle SU \rangle_t^{k-1-p}}{\langle S\rangle_t^{k-1}}\\[10pt]
    \times{k-1 \choose p}{p \choose n}[1-(1-\beta_A)^n(1-\beta_I)^{p-n}]\alpha_{EA}.
    \end{array}
    \label{eq:PiSE_chiusaPair}
\end{equation}
This expression can be further simplified. First, similarly to what we have done for the state probabilities $\langle EX\rangle$, we introduce the notation
\begin{equation}
    \sigma_A = \frac{\langle SA\rangle_t}{\langle S\rangle_t}, \qquad \sigma_I = \frac{\langle SI\rangle_t}{\langle S\rangle_t}, \qquad \sigma_U = \frac{\langle SU\rangle_t}{\langle S\rangle_t}.
\end{equation}
Then, we can manipulate Eq.~(\ref{eq:PiSE_chiusaPair}) so to have 
\begin{equation}
    \begin{array}{l}
    \Pi_{SE\rightarrow EA} \approx \langle SE \rangle_t [1 - \sum\limits_{p=0}^{k-1}\sum\limits_{n=0}^{p}\sigma_A^{n}\sigma_I^{p-n}\sigma_U^{k-1-p}\\[10pt]
    \times\frac{(k-1)!}{n!(p-n)!}(1-\beta_A)^n(1-\beta_I)^{p-n}]\alpha_{EA}\\[10pt]
    = \langle SE \rangle_t \{1 - \sum\limits_{p=0}^{k-1}\sum\limits_{n=0}^{p}\frac{(k-1)!}{n!(p-n)!(k-1-p)!}\sigma_U^{k-1-p}\\[10pt]
    \times[\sigma_A(1-\beta_A)]^n[\sigma_I(1-\beta_I)]^{p-n}\}\alpha_{EA} \\[10pt]
   = \langle SE \rangle_t [1 - \sum\limits_{p=0}^{k-1}\sum\limits_{n=0}^{p}\frac{(k-1)!}{n!(p-n)!(k-1-p)!}x_1^{k-1-p}x^{n}_{2} x_3^{p-n}]\alpha_{EA},
    \end{array}
    \label{eq:PiSE_chiusaPair_middle}
\end{equation}
where we have defined $x_1 = \sigma_U$, $x_2=\sigma_A(1-\beta_A)$ and $x_3=\sigma_I(1-\beta_I)$. Note also that we added to the summation the term relative to $p=0$, since this is equal to zero. By using the multinomial theorem \eqref{eq:multinomial_theorem} of power $k-1$, 
we can greatly simplify the expression of the transition probability as
\begin{equation}
    \begin{array}{l}
    \Pi_{SE\rightarrow EA} \approx \langle SE \rangle_t [1 - (x_1+x_2+x_3)^{k-1}]\alpha_{EA}.
    \end{array} 
    \label{eq:PiSE_chiusaPair_final}
\end{equation}
Finally, by substituting $x_1+x_2+x_3 = 1 - \beta_A\sigma_A - \beta_I\sigma_I$ (note that $\sigma_U = 1 - \sigma_A - \sigma_I$), and by returning to the usual notation for the state probabilities, we obtain for $\Pi_{SE\rightarrow EA}$ the expression in Eq.~(\ref{eq:pairwise_population_SE_EA}).

\section*{Appendix C: Pairwise population-level transition terms.}
\label{App_C}
\setcounter{equation}{0}
\renewcommand{\theequation}{C\arabic{equation}}
Here we present the complete list of the transition probability terms of the SEAIR pairwise model at the population-level. These terms can be derived in a way analogous to Eq.~(\ref{eq:transition_pair}), following the algebraic passages detailed in Appendix B. 
We begin by considering the nonlinear terms, and in particular the probability that a link in state $(S,S)$ transits to another state. i.e. $(E,E)$ or $(S,E)$. Let us first write an expression for the term $\Pi_{SS \rightarrow EE}$ in the form of Eq.~\eqref{eq:double_sum}. We have 
\begin{equation}
    \begin{array}{l}
    \Pi_{SS\rightarrow EE} \approx \langle SS \rangle_t \sum\limits_{p=1}^{k-1}\sum\limits_{n=0}^{p} \frac{\langle SA \rangle_t^{n}\langle SI \rangle_t^{p-n}\langle SU \rangle_t^{k-1-p}}{\langle S\rangle_t^{k-1}}\\[10pt]
    \times{k-1 \choose p}{p \choose n}[1-(1-\beta_A)^n(1-\beta_I)^{p-n}]\\[10pt] \times\sum\limits_{q=1}^{k-1}\sum\limits_{m=0}^{q} \frac{\langle SA \rangle_t^{m}\langle SI \rangle_t^{q-m}\langle SU \rangle_t^{k-1-q}}{\langle S\rangle_t^{k-1}}\\[10pt]
    \times{k-1 \choose q}{q \choose m}[1-(1-\beta_A)^m(1-\beta_I)^{q-m}],\\[10pt]
    \end{array}
    \label{eq:SS_EE_double_sum}
\end{equation}
where we note that the two double summation are equal. Similarly to what we have done for the expression of $\Pi_{SE \rightarrow EA}$, we can rewrite the previous equation as
\begin{equation}
    \begin{array}{l}
  \Pi_{SS \rightarrow EE} \approx \langle SS \rangle_t \left[1-\left(1-\beta_A \frac{\langle SA \rangle_t}{\langle S \rangle_t} - \beta_I \frac{\langle SI \rangle_t}{\langle S \rangle_t}\right)^{k-1}\right]^2,\\[10pt]
    \end{array}
    \label{eq:PiSS2EE}
\end{equation}
where the second power comes from the fact that both nodes undergo the same transition, i.e. $S\rightarrow E$. The term in the square bracket represents the probability that a node in state S is infected by at least one of its neighbors, which can be either in state A or in state I. Therefore, to write the expression of $\Pi_{SS \rightarrow SE}$, we have to consider the probability of the complementary event, namely the probability that none of the infected neighbors transmits the disease to the susceptible node. We thus have
\begin{equation}
    \begin{array}{l}
 \Pi_{SS \rightarrow SE} = \langle SS  \rangle_t \left[1-\left(1-\beta_A \frac{\langle SA \rangle_t}{\langle S \rangle_t} - \beta_I \frac{\langle SI \rangle_t}{\langle S\rangle_t}\right)^{k-1}\right]\\[10pt]
 \times\left(1-\beta_A \frac{\langle SA \rangle_t}{\langle S \rangle_t} - \beta_I \frac{\langle SI \rangle_t}{\langle S \rangle_t}\right)^{k-1}.\\[10pt]
    \end{array}
\end{equation}
Then, we account for all the possible transitions of links in state $(S,E)$. Starting from the expression for $\Pi_{SE \rightarrow EA}$ in Eq.~\eqref{eq:pairwise_population_SE_EA}, we can write the remaining terms by considering the probability of the complementary events, obtaining
\begin{equation}
    \begin{array}{l}
\Pi_{SE\rightarrow EE} \approx \\
\langle SE \rangle_t \left[1-\left(1-\beta_A \frac{\langle SA \rangle_t}{\langle S \rangle_t} - \beta_I \frac{\langle SI \rangle_t}{\langle S \rangle_t}\right)^{k-1}\right]\left(1-\alpha_{EA}\right),\\[10pt]
\Pi_{SE\rightarrow SA} \approx \\
\langle SE \rangle_t \left(1-\beta_A \frac{\langle SA \rangle_t}{\langle S \rangle_t} - \beta_I \frac{\langle SI \rangle_t}{\langle S \rangle_t}\right)^{k-1}\alpha_{EA}\\[10pt]
\end{array}
\label{eq:PiSE2EE}
\end{equation}
We now consider the possible transitions from state $(S,A)$. Let us take into account the probability term $\Pi_{SA \rightarrow EI}$ and let us come back for a moment to the expression with the double summation. This reads
\begin{equation}
\begin{array}{l}
\Pi_{SA\rightarrow EI} \approx \langle SA \rangle_t \sum\limits_{p=0}^{k-1} \sum\limits_{n=0}^{p}{k-1\choose p}{p\choose n} \frac{\langle SI \rangle_t^{p-n} \langle SA\rangle_t^{n} \langle SI\rangle_t^{k-1-n}}{\langle S \rangle_t^{k-1}}\\[10pt] 
\times\left[1-\left(1-\beta_A\right)^{n+1}\left(1-\beta_I\right)^{p-n}\right]\alpha_{AI},\\[10pt]
\end{array}
\end{equation}
where we have already written the probability that the node in state A transits to state I as $\alpha_{AI}$ (see Appendix B for more details). It is worth pointing out that, differently from the previous cases, the previous expression has a term $(1-\beta_A)$ raised to $(n+1)$-th power, which comes from the fact that, as we are considering the links in state $(S,A)$, the node in state S will always have an infectious neighbor. Following the same algebraic steps described in Appendix B, it is easy to verify that this expression can be simplified as
\begin{equation}
\begin{array}{l}
\Pi_{SA\rightarrow EI} \approx\\
\langle SA \rangle_t\left[1-\left(1-\beta_A \frac{\langle SA \rangle_t}{\langle S \rangle_t} - \beta_I \frac{\langle SI \rangle_t}{\langle S \rangle_t}\right)^{k-1}\left(1-\beta_A\right)\right]\alpha_{AI}.
\end{array}
\end{equation}
Once again, by considering the probability of the complementary events, we can write the remaining transition probabilities for the state $(S,A)$. Note that, since a node in state A can transit to two different states, i.e. to state I with probability $\alpha_{AI}$ and state R with probability $\mu_A$, the probability that the node remains in state A is given by $(1-\alpha_{AI}-\mu_A)$. We have
\begin{equation}
    \begin{array}{l}
\Pi_{SA\rightarrow ER} \approx \\
\langle SA \rangle_t \left[1-\left(1-\beta_A \frac{\langle SA \rangle_t}{\langle S \rangle_t} - \beta_I \frac{\langle SI \rangle_t}{\langle S \rangle_t}\right)^{k-1}\left(1-\beta_A\right)\right]\mu_A,\\[10pt]
\Pi_{SA\rightarrow EA} \approx \\
\langle SA \rangle_t \left[1-\left(1-\beta_A \frac{\langle SA \rangle_t}{\langle S \rangle_t} - \beta_I \frac{\langle SI \rangle_t}{\langle S \rangle_t}\right)^{k-1}\left(1-\beta_A\right)\right]\\[10pt]
\times\left(1-\alpha_{AI} -\mu_A\right),\\[10pt]
\Pi_{SA\rightarrow SI} \approx \\
\langle SA \rangle_t\left(1-\beta_A \frac{\langle SA \rangle_t}{\langle S \rangle_t} - \beta_I \frac{\langle SI \rangle_t}{\langle S \rangle_t}\right)^{k-1}\left(1-\beta_A\right)\alpha_{AI},\\[10pt]
\Pi_{SA\rightarrow SR} \approx \\
\langle SA \rangle_t\left(1-\beta_A \frac{\langle SA \rangle_t}{\langle S \rangle_t} - \beta_I \frac{\langle SI \rangle_t}{\langle S \rangle_t}\right)^{k-1}\left(1-\beta_A\right)\mu_{A}.\\[10pt]
\end{array}
\end{equation}
Analogously to the previous case, it is easy to write the transition probability terms from the state $(S,I)$. We have
\begin{equation}
    \begin{array}{l}
\Pi_{SI\rightarrow ER} \approx \\ 
\langle SI \rangle_t \left[1-\left(1-\beta_A \frac{\langle SA \rangle_t}{\langle S \rangle_t} - \beta_I \frac{\langle SI \rangle_t}{\langle S \rangle_t}\right)^{k-1}\left(1-\beta_I\right) \right]\mu_I,\\[10pt]
\Pi_{SI\rightarrow EI} \approx \\ 
\langle SI \rangle_t \left[1-\left(1-\beta_A \frac{\langle SA \rangle_t}{\langle S \rangle_t} - \beta_I \frac{\langle SI \rangle_t}{\langle S \rangle_t}\right)^{k-1}\left(1-\beta_I\right) \right]\left(1-\mu_I\right),\\[10pt]
\Pi_{SI\rightarrow SR} \approx \\ 
\langle SI \rangle_t \left(1-\beta_A \frac{\langle SA \rangle_t}{\langle S \rangle_t} - \beta_I \frac{\langle SI \rangle_t}{\langle S \rangle_t}\right)^{k-1}\left(1-\beta_I\right)\mu_I,\\[10pt]
\end{array}
\end{equation}
As concerns the state $(S,R)$, as R is an absorbing state, i.e. the node remains in R with probability equal to 1, the only possible transition is determined by the probability that the node in state S transits to state E, namely    
\begin{equation}
    \begin{array}{l}
\Pi_{SR\rightarrow ER} \approx \langle SR \rangle_t \left[1-\left(1-\beta_A \frac{\langle SA \rangle_t}{\langle S \rangle_t} - \beta_I \frac{\langle SI \rangle_t}{\langle S \rangle_t}\right)^{k-1}\right].
\end{array}
\end{equation}

Finally, we report the linear transition probability terms appearing in Eq.~\eqref{eq:pair}. All these terms can be expressed in the form of Eq.~\eqref{eq:linear_trans_pair}, so that they read:

\begin{equation}
    \begin{array}{lll}
\Pi_{EE\rightarrow AA} &=& \langle EE \rangle_t \alpha_{EA}^{2},\\
\Pi_{EE\rightarrow EA} &=& \langle EE \rangle_t (1-\alpha_{EA})\alpha_{EA},\\[5pt]

\Pi_{EA\rightarrow AI} &=& \langle EA \rangle_t \alpha_{EA}\alpha_{AI},\\
\Pi_{EA\rightarrow AR} &=& \langle EA \rangle_t \alpha_{EA}\mu_A,\\
\Pi_{EA\rightarrow AA} &=& \langle EA \rangle_t \alpha_{EA}(1-\mu_A-\alpha_{AI}),\\
\Pi_{EA\rightarrow EI} &=& \langle EA \rangle_t (1-\alpha_{EA})\alpha_{AI},\\
\Pi_{EA\rightarrow ER} &=& \langle EA \rangle_t (1-\alpha_{EA})\mu_{A},\\[5pt]

\Pi_{EI\rightarrow AR} &=& \langle EI \rangle_t \alpha_{EA} \mu_I,\\
\Pi_{EI\rightarrow AI} &=& \langle EI \rangle_t \alpha_{EA}(1-\mu_I),\\
\Pi_{EI\rightarrow ER} &=& \langle EI \rangle_t (1-\alpha_{EA})\mu_I,\\[5pt]

\Pi_{ER\rightarrow AR} &=& \langle ER \rangle_t \alpha_{EA},\\[10pt]

\Pi_{AA\rightarrow II} &=& \langle AA \rangle_t \alpha_{AI}^{2},\\
\Pi_{AA\rightarrow IR} &=& \langle AA \rangle_t \alpha_{AI}\mu_A,\\
\Pi_{AA\rightarrow RR} &=& \langle AA \rangle_t \mu_A^{2},\\
\Pi_{AA\rightarrow AI} &=& \langle AA \rangle_t (1-\mu_A-\alpha_{AI})\alpha_{AI},\\
\Pi_{AA\rightarrow AR} &=& \langle AA \rangle_t (1-\mu_A-\alpha_{AI})\mu_A,\\[5pt]

\Pi_{AI\rightarrow IR} &=& \langle AI \rangle_t [\alpha_{AI}\mu_I+\mu_{A}(1-\mu_I)],\\
\Pi_{AI\rightarrow RR} &=& \langle AI \rangle_t \mu_A\mu_I,\\
\Pi_{AI\rightarrow II} &=& \langle AI \rangle_t \alpha_{AI}(1-\mu_I),\\
\Pi_{AI\rightarrow AR} &=& \langle AI \rangle_t (1-\mu_A-\alpha_{AI})\mu_I,\\[5pt]

\Pi_{AR\rightarrow IR} &=& \langle AR \rangle_t \alpha_{AI},\\
\Pi_{AR\rightarrow RR} &=& \langle AR \rangle_t \mu_A,\\[5pt]

\Pi_{II\rightarrow RR} &=& \langle II \rangle_t \mu_I^2,\\
\Pi_{II\rightarrow IR} &=& \langle II \rangle_t (1-\mu_I)\mu_I,\\[5pt]

\Pi_{IR\rightarrow RR} &=& \langle IR \rangle_t \mu_I.
\end{array}
\end{equation}
Note that the probability $\Pi_{AI\rightarrow IR}$ consists of two terms as there are two possible ways in which a link in state $(A,I)$ can transits to state $(I,R)$. First, the node in state A can transits in state I (the asymptomatic infectious individual develops the symptoms) while the node in state I transits to state R (the symptomatic infectious individual recovers). Second, the node in state A can transits to state R (the asymptomatic infectious individual recovers) while the node in state I remains in it. In other words, a link in state $(A,I)$ can transits either to state $(I,R)$ or to state $(R,I)$, according to different probabilities. Coherently, a link can transit to the state $(I,R)$ coming from two different states, namely $(A,I)$ and $(I,A)$, according to the same probabilities, which justifies the use of a single transition probability term $\Pi_{AI\rightarrow IR}$ in Eq.~\eqref{eq:pair}.  

\section*{Appendix D: Derivation of $\mathcal{R}_0$}
\label{App_D1}
\setcounter{equation}{0}
\renewcommand{\theequation}{D\arabic{equation}}

In this Appendix, we show how to derive the expression of $\mathcal{R}_0$ for both the individual-based and the pairwise SEAIR model. To do so, we use the next-generation matrix (NGM) approach \cite{diekmann1990definition}, developed for discrete-time epidemic models \cite{allen2008basic}. To begin with, we briefly discuss the method, and then we apply it to the deterministic models introduced in our work. 

To to calculate the basic reproduction number, rather than the full system of master equations, the subsystem describing the evolution of the \emph{infected states} may be considered. Here, we follow the terminology used in \cite{diekmann1990definition} and indicate with the term 'infected states' those compartments that are infectious (A and I) or have been exposed to the disease (E). As a vanishing fraction of individuals in an infected state indicates that the infection dies out, $\mathcal{R}_0$ can be derived studying the condition under which the disease-free equilibrium (DFE), i.e. the equilibrium at which the fraction of individuals in an infected compartment is zero, becomes unstable. Note that this equilibrium always exists. Hence, in the context of the SEAIR models, instead of considering all the states, we will only focus on the variables involving the E, A and I compartments. Hereafter, we will generally denote with $\mathbf{X}(t)\in\mathbb{R}^{d}$ the vector containing the value of the subsystem dynamical variables at time $t$. To study the stability of the DFE, one can linearize the infected subsystem around it, writing the corresponding master equations as 
\begin{equation}
    \mathbf{X}(t+1) = (\mathcal{T}+\Sigma)\mathbf{X}(t),
\end{equation}
where $\mathcal{T}$ is called the \textit{transmission} matrix, as it accounts for the nonlinear probability terms, i.e. the disease transmission, while $\Sigma$ is the \textit{transition} one, which accounts for the linear transitions within the system. From the matrices $\mathcal{T}$ and $\Sigma$, the so-called next-generation matrix (NGM) \cite{diekmann1990definition} can be computed as
\begin{equation}
\label{eq:ngm}
    \mathcal{K} = \mathcal{T}(\mathds{1}_d - \Sigma)^{-1},
\end{equation}
where $\mathds{1}_d$ is the identity matrix of size $d$. The basic reproduction number can be calculated as the spectral radius of this matrix
\begin{equation}
\label{eq:R_0}
    \mathcal{R}_0 = \rho(\mathcal{T}(\mathds{1}_d - \Sigma)^{-1})=\rho(\mathcal{K}).
\end{equation}
In fact, it is possible to prove that if $\mathcal{R}_0<1$ the DFE is asymptotically stable, whereas it is unstable if $\mathcal{R}_0>1$ \cite{diekmann1990definition, li2002applications, diekmann2010construction, allen2008basic}. 

\subsection*{Individual-based model}
Here we show how to construct the NGM 
for the individual-based SEAIR model. In this case, to calculate the value of $\mathcal{R}_0$ we can focus on the dynamics of $\langle E\rangle_t$, $\langle A\rangle_t$ and $\langle I\rangle_t$, which are the variables representing the fraction of infected individuals within the population. Thus, we define $\mathbf{X}(t) = (\langle E\rangle_t,\langle A\rangle_t,\langle I\rangle_t)^T$. Linearizing around the DFE the equations relative to the infected subsystem in Eqs.~\eqref{eq:individual}, we obtain the following transmission and transition matrices: 
\begin{equation}
\mathcal{T} = \left[\begin{matrix}0 & \beta_{A} k & \beta_{I} k\\0 & 0 & 0\\0 & 0 & 0\end{matrix}\right],
\end{equation}

\noindent and
\begin{equation}
\Sigma = \left[\begin{matrix}1 - \alpha_{EA} & 0 & 0\\\alpha_{EA} & 1 - \alpha_{AI} - \mu_A  & 0\\0 & \alpha_{AI} & 1 - \mu_I\end{matrix}\right].
\end{equation}
Considering~\eqref{eq:ngm}, we finally derive that the NGM is given by
\begin{equation}
 \mathcal{K} = \left[\begin{matrix}\frac{k(\alpha_{AI} \beta_{I}+\mu_I\beta_A)}{\mu_I \left(\alpha_{AI} + \mu_A\right)} & \frac{k(\alpha_{AI} \beta_{I} + \mu_I\beta_A)}{\mu_I \left(\alpha_{AI} + \mu_A\right)} & \frac{\beta_{I} k}{\mu_I}\\0 & 0 & 0\\0 & 0 & 0\end{matrix}\right],  
\end{equation}
whose spectral radius $\rho(\mathcal{K})$ gives the value of $\mathcal{R}_0$ reported in Eq.~\eqref{eq:R0_individual}.

\subsection*{Pairwise model}

Here, we calculate the NGM for the pairwise SEAIR model. As mentioned above, we have to focus on all variables involving the E, A and I compartments, which in the pairwise model are the pair probabilities $\langle EX\rangle_t$, $\langle AX\rangle_t$ and $\langle IX\rangle_t$, with $X\in\Omega$. However, the subsystem of interest consists of a set of twelve master equations of Eqs.~\eqref{eq:pair}, which can be unfeasible to deal with. To reduce the number of equations, we can do the following consideration. Given the relation on the marginal probabilities \eqref{eq:marginal}, when the pair probabilities approach zero, the individual probabilities go to zero as well. In other words, if no link in the network has one infected node at one of its end, then no node in the network will be infected. Therefore, considering Eq.~\eqref{eq:marginal} and the expression of the transition probabilities, i.e. Eqs.~\eqref{eq:linear_trans_pair},~\eqref{eq:pairwise_population_SE_EA} and~(C1)-(C10), we can rewrite the system~\eqref{eq:pair} as

\begin{equation}
\begin{array}{lll}
    \langle S\rangle_{t+1} &=& \langle S\rangle_{t} - \langle S \rangle_t \pi_{S} \\[3pt]
    \langle E\rangle_{t+1} &=& \langle E\rangle_{t} + \langle S \rangle_t \pi_{S} - \langle E\rangle_{t}\alpha_{EA}\\[3pt]
    \langle A\rangle_{t+1} &=& \langle A\rangle_{t} + \langle E\rangle_{t}\alpha_{EA} - \langle A\rangle_{t}\alpha_{AI} - \langle A\rangle_{t}\mu_{A}\\[3pt]
    \langle I\rangle_{t+1} &=& \langle I\rangle_{t} + \langle A\rangle_{t}\alpha_{AI} - \langle I\rangle_{t}\mu_{I}\\[3pt]
    \langle R\rangle_{t+1} &=& \langle R\rangle_{t} + \langle A\rangle_{t}\mu_{A} + \langle I\rangle_{t}\mu_{I}\\[3pt]
    \langle SS\rangle_{t+1} &=& \langle SS\rangle_{t} - 2\langle SS \rangle_t \pi_{SS}(1-\pi_{SS}) - \langle SS\rangle_{t}\pi_{SS}^2\\[3pt]
    \langle SE\rangle_{t+1} &=& \langle SE\rangle_{t} + \langle SS \rangle_t \pi_{SS}(1-\pi_{SS}) - \langle SE\rangle_{t}\alpha_{EA}\\[3pt]
    &&- \langle SE\rangle_{t}\pi_{SS}(1-\alpha_{EA}) \\[3pt]
    \langle SA\rangle_{t+1} &=& \langle SA\rangle_{t} + \langle SE \rangle_t (1-\pi_{SS})\alpha_{EA} - \langle SA\rangle_{t}\pi_{SA}\\[3pt]
    &&- \langle SA\rangle_{t}(1-\pi_{SA})(\alpha_{AI}+\mu_A) \\[3pt]
    \langle SI\rangle_{t+1} &=& \langle SI\rangle_{t} + \langle SA \rangle_t (1-\pi_{SA})\alpha_{AI} - \langle SI\rangle_{t}\pi_{SI}\\[3pt]
    &&- \langle SI\rangle_{t}(1-\pi_{SI})\mu_I, \\[3pt]
\end{array}
\label{eq:compact_form}
\end{equation}
where, for the purpose of simplification, we have used the notation
\begin{equation}
    \begin{array}{lll}
    \pi_{S} &=& \left[1-\left(1-\beta_A \frac{\langle SA \rangle_t}{\langle S \rangle_t} - \beta_I \frac{\langle SI \rangle_t}{\langle S \rangle_t}\right)^{k}\right]\\[10pt]
    \pi_{SS} &=& \left[1-\left(1-\beta_A \frac{\langle SA \rangle_t}{\langle S \rangle_t} - \beta_I \frac{\langle SI \rangle_t}{\langle S \rangle_t}\right)^{k-1}\right]\\[10pt]
    \pi_{SA} &=& \left[1-\left(1-\beta_A \frac{\langle SA \rangle_t}{\langle S \rangle_t} - \beta_I \frac{\langle SI \rangle_t}{\langle S \rangle_t}\right)^{k-1}(1-\beta_A)\right]\\ [10pt]
    \pi_{SI} &=& \left[1-\left(1-\beta_A \frac{\langle SA \rangle_t}{\langle S \rangle_t} - \beta_I \frac{\langle SI \rangle_t}{\langle S \rangle_t}\right)^{k-1}(1-\beta_I)\right].\\
    \end{array}
    \label{eq:compact_transmission}
\end{equation}
Note that Eqs.~\eqref{eq:compact_form} represent a closed set of equations. As we are interested in analyzing the early stage of the epidemic, we can linearize the infected subsystem, which consists in the equations describing the dynamics of $\langle E\rangle_{t}$, $\langle A\rangle_{t}$, $\langle I\rangle_{t}$, $\langle SE\rangle_{t}$, $\langle SA\rangle_{t}$, $\langle SI\rangle_{t}$, around the DFE, characterized by $\langle S\rangle_{t+1} \approx \langle SS\rangle_{t+1} \approx 1$, while all the other variables approaches zero. 
We find
\begin{equation}
\begin{array}{lll}
    \langle E\rangle_{t+1} &\approx& \langle E\rangle_{t}(1-\alpha_{EA})\\[3pt] && + (k-1)(\beta_A\langle SA \rangle_t+\beta_I\langle SI \rangle_t)\\[3pt]
    \langle A\rangle_{t+1} &\approx& \langle E\rangle_{t}\alpha_{EA} + \langle A\rangle_{t}(1-\alpha_{AI} - \mu_{A})\\[3pt]
    \langle I\rangle_{t+1} &\approx& \langle A\rangle_{t}\alpha_{AI} + \langle I\rangle_{t}(1-\mu_{I})\\[3pt]
    \langle SE\rangle_{t+1} &\approx& \langle SE\rangle_{t}(1-\alpha_{EA})\\[3pt]&& + (k-1)(\beta_A\langle SA \rangle_t+\beta_I\langle SI \rangle_t)\\[3pt]
    \langle SA\rangle_{t+1} &\approx& \langle SA\rangle_{t}(1-\beta_A)(1-\alpha_{AI}-\mu_A) + \langle SE \rangle_t\alpha_{EA}\\[3pt]
    \langle SI\rangle_{t+1} &\approx& \langle SI\rangle_{t}(1-\beta_I)(1-\mu_I) + \langle SA \rangle_t (1-\beta_A)\alpha_{AI},\\[3pt]
\end{array}
\label{eq:compact_form_linearized}
\end{equation}
from which we can compute the transmission and transition matrices as
\begin{equation}
    \mathcal{T} = \left(\begin{matrix}0 & 0 & 0 & 0 & \beta_{A} k & \beta_{I} k\\0 & 0 & 0 & 0 & 0 & 0\\0 & 0 & 0 & 0 & 0 & 0\\0 & 0 & 0 & 0 & \beta_{A} \left(k - 1\right) & \beta_{I} \left(k - 1\right)\\0 & 0 & 0 & 0 & 0 & 0\\0 & 0 & 0 & 0 & 0 & 0\end{matrix}\right),
\end{equation}

\noindent and

\begin{widetext}
\begin{equation}
    \Sigma = \left(\begin{matrix}1 - \alpha_{EA} & 0 & 0 & 0 & 0 & 0\\\alpha_{EA} &1 - \alpha_{AI} - \mu_{A}  & 0 & 0 & 0 & 0\\0 & \alpha_{AI} & 1 - \mu_{I} & 0 & 0 & 0\\0 & 0 & 0 & 1 - \alpha_{EA} & 0 & 0\\0 & 0 & 0 & \alpha_{EA} & \left(1 - \beta_{A}\right) \left(1 - \alpha_{AI} - \mu_{A} \right) & 0\\0 & 0 & 0 & 0 & \alpha_{AI} \left(1 - \beta_{A}\right) & \left(1 - \beta_{I}\right) \left(1 - \mu_{I}\right)\end{matrix}\right).
\end{equation}
\end{widetext}
Finally, given the matrices $\mathcal{T}$ and $\Sigma$, we calculate the NGM matrix through \eqref{eq:ngm}, obtaining 
\begin{widetext}
\begin{equation}
\mathcal{K} = 
    \left(\begin{matrix}0 & 0 & 0 &  \frac{k \{\alpha_{AI} \beta_{I} \left(1 - \beta_{A}\right) + \beta_{A} \left[1- \left(1 - \beta_{I}\right) \left(1 - \mu_{I}\right)\right]\}}{\left[1- \left(1 - \beta_{A}\right) \left(1 - \alpha_{AI} - \mu_{A}\right) \right] \left[1- \left(1 - \beta_{I}\right) \left(1 - \mu_{I}\right) \right]} & \frac{k \{\alpha_{AI} \beta_{I} \left(1 - \beta_{A}\right) + \beta_{A} \left[1- \left(1 - \beta_{I}\right) \left(1 - \mu_{I}\right)\right]\}}{\left[1- \left(1 - \beta_{A}\right) \left(1 - \alpha_{AI} - \mu_{A}\right) \right] \left[1- \left(1 - \beta_{I}\right) \left(1 - \mu_{I}\right) \right]}& \frac{\beta_{I} k}{1 - \left(1- \beta_{I}\right)\left(1- \mu_{I}\right)}\\0 & 0 & 0 & 0 & 0 & 0\\0 & 0 & 0 & 0 & 0 & 0\\0 & 0 & 0 &  \frac{\left(k - 1\right) \{\alpha_{AI} \beta_{I} \left(1 - \beta_{A}\right) + \beta_{A} \left[1- \left(1 - \beta_{I}\right) \left(1 - \mu_{I}\right)\right]\}}{\left[1- \left(1 - \beta_{A}\right) \left(1 - \alpha_{AI} - \mu_{A}\right) \right] \left[1- \left(1 - \beta_{I}\right) \left(1 - \mu_{I}\right) \right]} & \frac{\left(k - 1\right) \{\alpha_{AI} \beta_{I} \left(1 - \beta_{A}\right) + \beta_{A} \left[1- \left(1 - \beta_{I}\right) \left(1 - \mu_{I}\right)\right]\}}{\left[1- \left(1 - \beta_{A}\right) \left(1 - \alpha_{AI} - \mu_{A}\right) \right] \left[1- \left(1 - \beta_{I}\right) \left(1 - \mu_{I}\right) \right]} & \frac{\beta_{I} \left(k-1\right)}{1 - \left(1- \beta_{I}\right)\left(1- \mu_{I}\right)}\\0 & 0 & 0 & 0 & 0 & 0\\0 & 0 & 0 & 0 & 0 & 0\end{matrix}\right),
\end{equation}
\end{widetext}

The last step is to compute the spectral radius $\rho(\mathcal(K))$ that gives the expression of the basic reproduction number $\mathcal{R}_0$ for the pairwise SEAIR model reported in Eq.~\eqref{eq:R0_pair}.

\bibliography{pairwise.bib}

\end{document}